\documentclass[a4paper,fleqn,usenatbib]{mnras}

\usepackage[T1]{fontenc}
\usepackage{ae,aecompl}

\usepackage{amssymb}	
\usepackage[font=small,labelfont=bf]{caption}
\usepackage{amssymb,amsmath}
\usepackage{afterpage} 
\usepackage{amsmath} 
\usepackage{booktabs} 
\usepackage{enumerate} 
\usepackage{rotating} 
\usepackage{graphicx} 
\usepackage{subcaption}
\usepackage{hyperref} 
\usepackage{url}
\usepackage{longtable}
\usepackage{makeidx}
\usepackage{url}
\usepackage{natbib}
\usepackage{multirow}
\usepackage{epstopdf}
\usepackage{caption}
\usepackage[section]{placeins}
\usepackage{setspace}
\DeclareCaptionFormat{cont}{#1 (cont.)#2#3\par}
\graphicspath{ {figures/} }

\title[NIR counterparts of ULXs]{A Systematic Search for Near-Infrared Counterparts of Nearby Ultraluminous X-ray sources (II)}

\author[K. M. L\'{o}pez et al.]{
K. M. L\'{o}pez$^{1,2}$\thanks{Contact e-mail: \href{mailto:K.M.Lopez@sron.nl}{K.M.Lopez@sron.nl}},
M. Heida$^{3}$,
P. G. Jonker$^{1,2}$,
M. A. P. Torres$^{1,2}$
\newauthor
T. P. Roberts$^{4}$,
D. J. Walton$^{3}$,
D.-S. Moon$^{5}$,
F. A. Harrison$^{3}$
\\
\\
$^{1}$SRON Netherlands Institute for Space Research, 3584 CA Utrecht, The Netherlands
\\
$^{2}$Department of Astrophysics/IMAPP, Radboud University, P.O. Box 9010, 6500 GL Nijmegen, The Netherlands
\\
$^{3}$Space Radiation Laboratory, California Institute of Technology, Pasadena, CA 91125, USA
\\
$^{4}$Centre for Extragalactic Astronomy, Department of Physics, University of Durham, South Road, Durham DH1 3LE, United Kingdom
\\
$^{5}$Department of Astronomy and Astrophysics, University of Toronto, Toronto, ON M5S 3H4, Canada
}

\date{Accepted 2017 April 3. Received 2017 April 3; in original form 2017 February 1}

\pubyear{2016}

\begin{document}
\label{firstpage}
\pagerange{\pageref{firstpage}--\pageref{lastpage}}
\maketitle

\begin{abstract}
We present the results of our continued systematic search for near-infrared (NIR)
candidate counterparts to ultraluminous X-ray sources (ULXs) within 10 Mpc.
We observed 42 ULXs in 24 nearby galaxies and detected NIR candidate counterparts
to 15 ULXs. Fourteen of these ULXs appear to have a single candidate counterpart
in our images and the remaining ULX has 2 candidate counterparts.
Seven ULXs have candidate counterparts with
absolute magnitudes in the range between -9.26 and -11.18 mag, consistent with them
being red supergiants (RSGs). The other eight ULXs have candidate counterparts with 
absolute magnitudes too bright to be a single stellar source. 
Some of these NIR sources show extended morphology
or colours expected for Active Galactic Nuclei (AGN), strongly suggesting that they are likely
stellar clusters or background galaxies. 
The red supergiant candidate counterparts form a valuable
sample for follow-up spectroscopic observations to confirm their nature, with the
ultimate goal of directly measuring the mass of the compact accretor that powers the ULX 
using binary Doppler shifts.
\end{abstract}




\section{Introduction}
\label{intro}

Ultraluminous X-ray sources (ULXs) are defined as point-like, off-nuclear 
sources with an X-ray luminosity that exceeds 10$^{39}$ erg/s
\citep{2011NewAR..55..166F}, the Eddington luminosity of a 10 M$_{\odot}$ black hole 
\citep{2006tmgm.meet..530C}. These high luminosities have been explained
in different ways. In a first scenario, a ULX could be a stellar mass
black hole emitting anisotropically
\citep{2001ApJ...552L.109K}, or alternatively, at super-Eddington luminosity 
\citep{2002ApJ...568L..97B,2003ApJ...586.1280M,2009MNRAS.397.1836G}. Examples of
these are the neutron star ULXs 
\citep{2014Natur.514..202B,2016arXiv160907375I,2017MNRAS.466L..48I}. 
Another possibility is
that the black hole is more massive than the typical 10 M$_{\odot}$ 
stellar remnants seen in our own Galaxy, 
resulting in a higher Eddington luminosity. A different hypothesis is that some ULXs
could be recoiling supermassive black holes (SMBHs) whereas some could instead be
remnant black holes from a smaller galaxy that underwent a merger with a larger
galaxy \citep{2010MNRAS.407..645J}.
The latter scenarios assume
that ULXs host massive (10$^2$-10$^5$ M$_{\odot}$) black holes.
We call these intermediate mass black holes (IMBHs). 
The recent detection of gravitational waves from a binary 
black hole merger,  where the estimated mass of the final black hole is 62 
M$_{\odot}$ \citep{2016PhRvL.116f1102A}, proves the existence of more massive 
black holes than stellar mass black holes 
previously observed \citep{2007Natur.449..872O,2014SSRv..183..223C}. 
IMBHs could be the building blocks
of SMBHs \citep{2001ApJ...562L..19E} and if they exist they could help explain
the puzzling observation that even at $z$ > 6, SMBHs with masses > 10$^9$ 
M$_{\odot}$ exist \citep{2000AJ....120.1167F,2015Natur.518..512W,2016arXiv160803279B}. 
In some sources, the IMBH 
interpretation of ULX is supported by luminosities $\gtrsim$ 10$^{41}$ erg/s, 
as such are difficult to achieve with current models of highly super-Eddington
accretion \citep{2012MNRAS.423.1154S}.

However, there exists no definitive evidence for an IMBH yet. The most 
reliable way to determine the true nature of ULXs is to determine
the mass of their accretors via a dynamical mass measurement. To date,
the most reliable mass constraint is that of the neutron star ULX M82-X2 
\citep{2014Natur.514..202B},  estimated through the detection of pulsations in this
source and the inference that the upper limit to the mass of a neutron star is 3 M$_{\odot}$ \citep{1996ApJ...470L..61K}.

Several studies have focused on detecting optical counterparts
to ULXs (e.g. \citealt{2006ApJS..166..154P,2006IAUS..230..310G,2011AN....332..398R,2013ApJS..206...14G,2015NatPh..11..551F}) 
and on radio counterparts (e.g. \citealt{2011AN....332..384P}). 
Since the optical counterparts are often faint (V > 24 mag),
radial velocity studies have made use of emission lines.
However, these attempts have encountered difficulties, as the emission lines originate
in the accretion disk and/or the surrounding nebulae, 
not the companion star itself. 

Others have focused on absorption lines from the donor stars
on the blue part of the spectrum (e.g. \citealt{2014Natur.514..198M}), as the observed colours are consistent with that of blue mass donors. In addition, several ULXs are located
in or near young star clusters 
(e.g. \citealt{2001ApJ...554.1035F,2002MNRAS.337..677R,2003ApJ...596L.171G,2013MNRAS.432..506P}), 
and thus, a blue early-type donor star might be expected certainly for those ULXs (e.g. \citealt{2012ApJ...758...28J}).

The association
with young star clusters implies that some of the donor stars can be red supergiants (RSGs, 
\citealt{2005MNRAS.362...79C,2007MNRAS.376.1407C,2008MNRAS.386..543P,2014MNRAS.442.1054H}), 
which are very bright in the near-infrared (NIR) band. Therefore, 
\citet{2014MNRAS.442.1054H} (hereafter H14)
performed the first systematic search for
NIR counterparts to nearby (D < 10 Mpc) ULXs. Observing 62 ULXs, they discovered 
17 candidate NIR counterparts, 11 of which had an absolute magnitude consistent 
with that of a RSG (see Table~\ref{tab:cand-marianne}). 
During initial spectroscopic follow-up,
they discovered RSG counterparts to ULX RX J004722.4-252051 
(in NGC 253, \citealt{2015MNRAS.453.3510H}), ULX J022721+333500 (in NGC 925) and ULX 
J120922+295559 (in NGC 4136, \citealt{2016MNRAS.459..771H}).

In this paper we present the results of our continued systematic search for
candidate RSG counterparts to ULXs within 10 Mpc from our Galaxy. We describe
the sample in Section~\ref{Sample}, the NIR observations and data 
reduction/photometry in Section~\ref{nirobs}. The X-ray astrometric correction 
is explained in Section~\ref{xraydata} and our results are presented and 
discussed in detail in Section~\ref{results}. We end
with the conclusions of our work in Section~\ref{conclusions}.

\section{Sample}
\label{Sample}

Our sample consists of 45\footnote{We took 45 ULXs from the catalogues below, but we revise the number to 42 as in two cases 2 entries are likely from the same source (see subsection~\ref{rosat}), and one source is likely not a ULX (see subsection~\ref{ulx7}).} ULXs located in 24 galaxies within 10 Mpc from our
own Galaxy (see Table~\ref{tab:galaxies}), since this is the maximum distance at which 
it is possible to take NIR spectra of a RSG with existing telescopes (H14). 
This imaging campaign
almost completes the ULX sample within 10 Mpc taken from the catalogues
of \citet{2005ApJS..157...59L,2005A&A...429.1125L,2006ApJ...649..730W,2004ApJS..154..519S,2011ApJ...741...49S,2011ApJS..192...10L,
2011MNRAS.416.1844W} and Earnshaw et al.
(in prep.). Six ULXs were observed before by H14, but
we observed them again under better sky conditions and in the $H$-band instead of 
the $K$-band in 5 of the 6 cases.

\section{NIR Observations}
\label{nirobs}

$H$-band imaging of regions of galaxies containing ULXs were obtained with the 
Long-slit Intermediate Resolution
Infrared Spectrograph (LIRIS) mounted on the William Herschel Telescope. 
LIRIS has a field of view of 
4.27\arcmin\  $\times$ 4.27\arcmin\  and a pixel scale of 0.25\arcsec\ /pixel. The observations were
performed using 7 or 8 repetitions of a 5-point dither pattern where 5
images (20 seconds exposure per image) were taken at each point.
Of the 24 galaxies, 8 were observed in April 2015, 9  in January
2016 and 9\footnote{Two galaxies from April 2015 were re-observed in January 2016.} 
in March 2016 (see Table~\ref{tab:galaxies} where  
the average seeing is provided, as a measure of the image quality during 
the observations).

\subsection{Data reduction}
\label{datared}

The data reduction was performed using the {\scshape theli} pipeline \citep{2013ApJS..209...21S}.
With {\scshape theli} we produced a master flat to flat-field correct the data and
we generate a sky background model which is subsequently subtracted from the individual
data frames.
In order to detect sources in the images and 
to obtain astrometric solutions {\scshape theli} uses {\scshape SExtractor}
\citep{1996A&AS..117..393B} and {\scshape scamp} \citep{2006ASPC..351..112B}, respectively.
The astrometric solution is obtained by matching the detected
positions to sources from the 2 Micron All Sky Survey
(2MASS; \citealt{2006AJ....131.1163S}) or PPXML
(Position and Proper Motion Extended-L; \citealt{2010AJ....139.2440R}).
The global astrometric solution is subsequently used for the coaddition of all the
images using {\scshape swarp} \citep{2002ASPC..281..228B}.

In order to obtain accurate astrometric positions, 
we improved the accuracy of the global astrometric solution 
of the coadded images using the {\scshape starlink}
tool {\scshape gaia}, fitting at least 5 star positions from the fourth US 
Naval Observatory CCD Astrograph Catalog (UCAC4, \citealt{2013AJ....145...44Z})
or 2MASS (if the field of view did not have 5 sources in UCAC4) to build a local astrometric
solution around the position of the ULX.
The rms errors of the fits are listed in Table~\ref{tab:galaxies}, 
indicated as WCS (World Coordinate System) uncertainties,
where the intrinsic error of the catalog with respect 
to the International Celestial Reference System (ICRS) is also indicated:
15 mas (systematic) for 2MASS and 20 mas (systematic) for UCAC4.
We were not able to improve the astrometry for one galaxy, NGC 4258 (observed
on March 26, 2016), since there were not enough reference stars in the vicinity (1\arcmin\ ) of the ULX.
For this galaxy, we indicate the uncertainty for the global astrometric solution
that {\scshape theli} provides.

\begin{table*}
\vspace{5mm}
\begin{center}
\caption{Galaxies observed in the $H$-band with the LIRIS instrument on the WHT.}
\label{tab:galaxies}
\resizebox{\textwidth}{!}{\begin{tabular}{|lcccccccc|}
\hline\hline
Galaxy & Date & Exposure & WCS & Zero point$^c$ & Limiting$^d$ & Average & Distance & Distance\\
 & observed & time$^a$ & uncertainty$^b$ & magnitude & magnitude & seeing & & Ref.\\
 & & (sec) & (mas) & (mag) & (mag) & ( \arcsec\ ) & (Mpc) & \\
\hline\hline
NGC 4190 & Apr 05, 2015 & 3920 & 253 $\pm$ 20 & 23.03 $\pm$ 0.08 & 19.06 $\pm$ 0.01 & 1.3 & 2.83 $\pm$ 0.28 & A\\
NGC 4559 & Apr 05, 2015 & 3060 & 203 $\pm$ 20 & 23.06 $\pm$ 0.10 & 18.73 $\pm$ 0.22 & 1.2 & 7.31 $\pm$ 1.46 & A\\
NGC 5194 & Apr 05, 2015 & 1720 & 164 $\pm$ 20 & 23.09 $\pm$ 0.09 & 18.82 $\pm$ 0.13 & 1.1 & 9.05 $\pm$ 0.24 & B\\
NGC 4490 & Apr 08, 2015 & 2580 & 120 $\pm$ 15 & 22.36 $\pm$ 0.13 & 18.48 $\pm$ 0.62 & 0.8 & 7.80 $\pm$ 0.62 & C\\
NGC 4485 & Apr 08, 2015 & 900 & 9.95 $\pm$ 15 & 22.61 $\pm$ 0.06 & 18.62 $\pm$ 0.28 & 0.7 & 8.91 $\pm$ 0.89 & A\\
NGC 4625 & Apr 08, 2015 & 4000 & 447 $\pm$ 15 & 23.22 $\pm$ 0.05 & 19.65 $\pm$ 0.05 & 1.0 & 8.20 $\pm$ 0.66 & C\\
NGC 4736 & Apr 08, 2015 & 1400 & 225 $\pm$ 15 & 22.27 $\pm$ 0.26 & 18.31 $\pm$ 0.04 & 0.8 & 4.59 $\pm$ 0.37 & A\\
NGC 5457 & Apr 08, 2015 & 1600 & 317 $\pm$ 15 & 23.20 $\pm$ 0.08 & 19.92 $\pm$ 0.28 & 0.8 & 6.95 $\pm$ 0.42 & A\\
NGC 891 & Jan 25, 2016 & 2400 & 109 $\pm$ 20 & 23.27 $\pm$ 0.03 & 19.11 $\pm$ 0.51 & 1.0 & 9.12 $\pm$ 0.73 & A\\
NGC 891 & Jan 25, 2016 & 3720 & 154 $\pm$ 20 & 23.32 $\pm$ 0.02 & 20.06 $\pm$ 0.36 & 1.0 & 9.12 $\pm$ 0.73 & A\\
NGC 2403 & Jan 25, 2016 & 4000 & 251 $\pm$ 20 & 23.07 $\pm$ 0.03 & 18.08 $\pm$ 0.22 & 1.1 & 3.18 $\pm$ 0.19 & A\\
NGC 3486 & Jan 25, 2016 & 3900 & 26.7 $\pm$ 15 & 22.96 $\pm$ 0.09 & 19.03 $\pm$ 0.27 & 1.2 & 7.40 $\pm$ 0.59 & C\\
NGC 1042 & Jan 26, 2016 & 3500 & 168 $\pm$ 15 & 23.27 $\pm$ 0.03 & 19.55 $\pm$ 0.35 & 0.9 & 4.21 $\pm$ 0.30 & D\\
NGC 2500 & Jan 26, 2016 & 3680 & 172 $\pm$ 15 & 23.40 $\pm$ 0.05 & 20.16 $\pm$ 0.06 & 0.8 & 10.10 $\pm$ 0.81 & C\\
NGC 2903 & Jan 26, 2016 & 4200 & 319 $\pm$ 15 & 23.11 $\pm$ 0.19 & 20.17 $\pm$ 0.07 & 0.7 & 9.46 $\pm$ 1.89 & A\\
NGC 3990 & Jan 26, 2016 & 3000 & 157 $\pm$ 20 & 23.34 $\pm$ 0.06 & 19.11 $\pm$ 0.19 & 0.9 & 10.05 $\pm$ 1.41 & A\\
IC 342 & Jan 27, 2016 & 3060 & 105 $\pm$ 20 & 22.99 $\pm$ 0.02 & 19.31 $\pm$ 0.41 & 0.7 & 2.73 $\pm$ 0.19 & A\\
NGC 855 & Jan 27, 2016 & 3660 & 9.3 $\pm$ 20 & 21.34 $\pm$ 0.01 & 18.37 $\pm$ 0.12 & 0.8 & 8.83 $\pm$ 1.24 & A\\
NGC 3031 & Mar 26, 2016 & 3960 & 317 $\pm$ 15 & 23.63 $\pm$ 0.03 & 18.94 $\pm$ 0.04 & 0.7 & 3.61 $\pm$ 0.22 & A\\
NGC 4594 & Mar 26, 2016 & 1600 & 170 $\pm$ 15 & 23.49 $\pm$ 0.03 & 19.99 $\pm$ 0.09 & 0.9 & 11.27 $\pm$ 1.35 & A\\
NGC 4258 & Mar 26, 2016 & 3020 & 829 $\pm$ 15 & 23.49 $\pm$ 0.03 & 20.03 $\pm$ 0.27 & 0.8 & 7.31 $\pm$ 0.37 & A\\
NGC 4258 & Mar 26, 2016 & 3500 & 263 $\pm$ 15 & 23.49 $\pm$ 0.03 & 20.55 $\pm$ 0.05 & 1.0 & 7.31 $\pm$ 0.37 & A\\
NGC 4631 & Mar 27, 2016 & 2800 & 497 $\pm$ 15 & 22.90 $\pm$ 0.01 & 18.53 $\pm$ 0.27 & 0.9 & 7.35 $\pm$ 0.74 & A\\
NGC 5128 & Mar 27, 2016 & 3500 & 80 $\pm$ 20 & 22.90 $\pm$ 0.01 & 17.79 $\pm$ 0.11 & 1.2 & 3.66 $\pm$ 0.22 & A\\
NGC 4517 & Mar 27, 2016 & 3480 & 180 $\pm$ 15 & 23.42 $\pm$ 0.06 & 19.19 $\pm$ 0.51 & 0.8 & 8.58 $\pm$ 0.77 & A\\
NGC 3521 & Mar 28, 2016 & 4000 & 107 $\pm$ 20 & 23.51 $\pm$ 0.04 & 20.17 $\pm$ 0.18 & 0.9 & 14.19 $\pm$ 2.84 & A\\
NGC 5194 & Mar 28, 2016 & 3600 & 147 $\pm$ 20 & 23.29 $\pm$ 0.04 & 18.09 $\pm$ 0.01 & 1.0 & 9.05 $\pm$ 0.24 & B\\
NGC 5457 & Mar 28, 2016 & 3500 & 38.8 $\pm$ 15 & 23.74 $\pm$ 0.06 & 20.35 $\pm$ 0.10 & 0.9 & 6.95 $\pm$ 0.42 & A\\
NGC 5457 & Mar 28, 2016 & 2920 & 225 $\pm$ 15 & 23.24 $\pm$ 0.06 & 20.47 $\pm$ 0.17 & 0.9 & 6.95 $\pm$ 0.42 & A\\
\hline\hline
\multicolumn{9}{l}{{\bf Notes:} $^a$:Effective exposure time, not necessarily equal to the net exposure time (i.e., images taken during twilight, cloudy}\\
\multicolumn{9}{l}{ intervals). $^b$:Uncertainty with respect to the reference catalog. The first value is a statistical uncertainty, given by the {\scshape gaia} tool;}\\
\multicolumn{9}{l}{while the second value is a systematic uncertainty, and corresponds to the astrometic accuracy of the reference catalog: 20 mas}\\
\multicolumn{9}{l}{for UCAC4 and 15 mas for 2MASS. $^c$:Zero point magnitude derived from the data, with respect to the 2MASS catalog. $^d$:See}\\
\multicolumn{9}{l}{text in section~\ref{Photometry} for how the limiting magnitude is defined. References: A: \citet{2013AJ....146...86T} B: \citet{2015AstL...41..239T}}\\
\multicolumn{9}{l}{C: \citet{1988ngc..book.....T} D: \citet{2009AJ....138..323T}.}\\
\end{tabular}}
\end{center}
\end{table*}

\subsection{Photometry}
\label{Photometry}

We used {\scshape SExtractor} for the source detection and photometry
in every NIR image, making sure that each detection was more than 3-$\sigma$ above 
the local background. We performed aperture photometry to
determine instrumental magnitudes.
As aperture size, we use the
average full-width at half maximum (FWHM) of the light distribution of 
point-like objects in each image determined with the
{\scshape starlink} tool {\scshape gaia}. 
The photometric zero points of our NIR images were measured by using isolated
2MASS objects in the field of view. These are given in Table~\ref{tab:galaxies}.
Instrumental magnitudes for all detected sources were converted to apparent magnitudes.
Finally, we determined the absolute magnitudes of any candidate counterparts to ULXs
taking the distances given in Table~\ref{tab:galaxies}.
The (1-$\sigma$) uncertainties on these values are estimated taking
into account the distance uncertainty, the uncertainty
in the determination of the zero point magnitude and the uncertainty 
in the instrumental magnitude given by {\scshape SExtractor}.

We estimated the limiting magnitude of each NIR image. To do this,
we made a histogram of the magnitudes of the detected objects
and the peak value of that histogram is a measure of the completeness limit, which
we conservatively take as a measure of the limiting magnitude. 
We fit a linear model to the bright source end of the histogram of the magnitudes 
of each NIR image and compare the peak value with the value given by the linear fit
to estimate the uncertainty on the limiting magnitude (see Table~\ref{tab:galaxies}).

\section{X-ray astrometric localization}
\label{xraydata}

For fourteen ULXs out of the 45 accurate positions existed in the literature (e.g. 
\citealt{2011ApJS..192...10L}, \citealt{2011ApJ...741...49S}, \citealt{2014MNRAS.442.1054H}). 
In addition, we were able to determine accurate positions for 19 sources using archival
{\it Chandra}/ACIS observations and for 9 sources using archival {\it XMM-Newton}
observations. The observation IDs and the exposure times are detailed
in Table~\ref{tab:corrected}. We explain the detailed analysis for these 28 sources
in sections~\ref{chandra} and~\ref{xmm}. 

For the remaining 3 ULXs, observed only by {\it ROSAT} High Resolution Imager
(HRI), we were not able to determine a position with higher 
accuracy than the original one from \citet{2005ApJS..157...59L}.

\begin{table*}
\vspace{5mm}
\begin{center}
\caption{ULXs for which we reduced the uncertainty on the localization, either through applying a boresight correction to the {\it Chandra} data, correcting the {\it XMM-Newton} position with the SAS software or improving the {\it ROSAT} position identifying the ULX in a {\it Chandra} image.}
\label{tab:corrected}
\resizebox{\textwidth}{!}{\begin{tabular}{|lcccccccccccc|}
\hline\hline
Galaxy & ULX ID & Observation & Exposure & Satellite & Catalog$^a$ & \multicolumn{7}{|c|}{Uncertainties}\\
 & as in & ID & time & & & {\scshape wavdetect}$^b$ & Boresight$^c$ & {\scshape catcorr}$^d$ & {\it Chandra} & \multicolumn{2}{c}{Catalog} & Total\\
 & Table~\ref{tab:ulxs} & & (ks) & & & ( \arcsec\ )$^e$ & ( \arcsec\ )$^e$ & ( \arcsec\ )$^f$  & ( \arcsec\ )$^f$ & ( \arcsec\ )$^e$ & ( \arcsec\ )$^f$ & ( \arcsec\ )$^g$\\
\hline\hline
NGC 891 & 2 & 794 & 50.9 & {\it Chandra} & 2MASS & 0.06 & 0.50 & - & - & - & 0.045 & 0.7\\
NGC 891 & 4 & 794 & 50.9 & {\it Chandra} & 2MASS & 0.06 & 0.50 & - & - & - & 0.045 & 0.7\\
NGC 1042 & 6 & 553300401 & 59.1 & {\it XMM-Newton} & SDSS9 & - & - & 0.70 & - & 0.18 & 0.09 & 1.4\\
IC 342 & 7 & 693851301 & 60.1 & {\it XMM-Newton} & SDSS9 & - & - & 1.13 & - & 0.18 & 0.09 & 1.6\\
NGC 2500 & 10 & 7112 & 2.5 & {\it Chandra} & SDSS7 & 0.31 & 0.74 & - & - & 0.18 & 0.09 & 1.1\\
NGC 2903 & 11 & 11260 & 94.8 & {\it Chandra} & - & 0.12 & - & - & 0.8 & - & - & 1.6\\
NGC 3990 & 15 & 90020101 & 13.2 & {\it XMM-Newton} & SDSS9 & - & - & 0.82 & - & 0.18 & 0.09 & 1.5\\
NGC 4258 & 17 & 59140901 & 16.5 & {\it XMM-Newton} & SDSS9 & - & - & 0.75 & - & 0.18 & 0.09 & 3.4\\
NGC 4258 & 18 & 400560301 & 64.5 & {\it XMM-Newton} & SDSS9 & - & - & 0.87 & - & 0.18 & 0.09 & 1.8\\
NGC 4490 & 19 & 1579 & 19.5 & {\it Chandra} & SDSS7 & 0.13 & 0.58 & - & - & 0.18 & 0.09 & 0.9\\
NGC 4485 & 20 & 1579 & 19.5 & {\it Chandra} & SDSS7 & 0.08 & 0.58 & - & - & 0.18 & 0.09 & 0.8\\
NGC 4490 & 21 & 1579 & 19.5 & {\it Chandra} & SDSS7 & 0.08 & 0.58 & - & - & 0.18 & 0.09 & 0.9\\
NGC 4490 & 22 & 1579 & 19.5 & {\it Chandra} & SDSS7 & 0.09 & 0.58 & - & - & 0.18 & 0.09 & 0.9\\
NGC 4490 & 23 & 1579 & 19.5 & {\it Chandra} & SDSS7 & 0.08 & 0.58 & - & - & 0.18 & 0.09 & 0.9\\
NGC 4490 & 24 & 1579 & 19.5 & {\it Chandra} & SDSS7 & 0.88 & 0.58 & - & - & 0.18 & 0.09 & 1.3\\
NGC 4490 & 25 & 1579 & 19.5 & {\it Chandra} & SDSS7 & 0.08 & 0.58 & - & - & 0.18 & 0.09 & 0.9\\
NGC 4517 & 26 & 203170301 & 114.3 & {\it XMM-Newton} & SDSS9 & - & - & 0.64 & - & 0.18 & 0.09 & 1.4\\
NGC 4594 & 29 & 84030101 & 43.5 & {\it XMM-Newton} & SDSS9 & - & - & 1.2 & - & 0.18 & 0.09 & 1.9\\
NGC 4625 & 30 & 9549 & 55.5 & {\it Chandra} & SDSS7 & 0.34 & 0.58 & - & - & 0.18 & 0.09 & 1.7\\
NGC 4631 & 31 & 797 & 59.2 & {\it Chandra} & SDSS7 & 0.20 & 0.77 & - & - & 0.18 & 0.09 & 1.8\\
NGC 4631 & 33 & 797 & 59.2 & {\it Chandra} & SDSS7 & 0.03 & 0.77 & - & - & 0.18 & 0.09 & 1.8\\
NGC 4631 & 34 & 797 & 59.2 & {\it Chandra} & SDSS7 & 0.11 & 0.77 & - & - & 0.18 & 0.09 & 1.8\\
NGC 4631 & 35 & 797 & 59.2 & {\it Chandra} & SDSS7 & 0.09 & 0.77 & - & - & 0.18 & 0.09 & 1.8\\
NGC 5128 & 39 & 724060701 & 26.8 & {\it XMM-Newton} & USNO-B1.0 & - & - & 1.36 & - & - & 0.6 & 2.3\\
NGC 5194 & 40 & 13814 & 189.8 & {\it Chandra} & - & 0.17 & - & - & 0.8 & - & - & 1.4\\
NGC 5194 & 41 & 13814 & 189.8 & {\it Chandra} & - & 0.11 & - & - & 0.8 & - & - & 1.3\\
NGC 5457 & 42 & 104260101 & 43.4 & {\it XMM-Newton} & USNO-B1.0 & - & - & 1.7 & - & - & 0.6 & 2.5\\
NGC 5457 & 45 & 2779 & 14.2 & {\it Chandra} & - & 0.24 & - & - & 0.8 & - & - & 1.9\\
\hline\hline
\multicolumn{13}{l}{{\bf Notes:} $^a$:Catalog used to perform the astrometric correction. $^b$:3-$\sigma$ uncertainty in the source localization on the CCDs given by the {\scshape wavdetect} task. $^c$:3-$\sigma$ }\\
\multicolumn{13}{l}{uncertainty obtained after performing the boresight correction. $^d$:3-$\sigma$ uncertainty after performing the astrometric correction for {\it XMM-Newton} data. $^e$: Statistical}\\
\multicolumn{13}{l}{uncertainty. $^f$: Systematic uncertainty of the catalogue used, (i.e. 0.8 \arcsec\ for {\it Chandra}, \href{http://cxc.harvard.edu/cal/ASPECT/celmon/}{Chandra X-ray Center}). $^g$:Calculated taking also into account the WCS}\\
\multicolumn{13}{l}{uncertainty from the NIR image (see Table~\ref{tab:galaxies}).}\\
\end{tabular}}
\end{center}
\end{table*}

\begin{table*}
\vspace{5mm}
\begin{center}
\caption{Complete sample of observed ULXs.}
\label{tab:ulxs}
\resizebox{\textwidth}{!}{\begin{tabular}{|cllcccccc|}
\hline\hline
ULX ID & Galaxy & ULX name & R.A. & Dec & 3-$\sigma$ position & Position & Telescope\\
 & & (SIMBAD$^{\ddagger}$) & & & uncertainty$^a$ & reference$^b$ &\\
 & & & (hh:mm:ss) & (dd:mm:ss) & (arcsec) & &\\ 
\hline\hline
1 & NGC 855$^*$ & [SST2011] J021404.08+275239.5 & 02:14:04.1 & 27:52:39.4 & 1.6 & \citet{2011ApJS..192...10L} & {\it Chandra}\\ 
2 & NGC 891 & [SST2011] J022231.26+421957.8 & 02:22:31.3 & 42:19:57.4 & 0.7 & This work & {\it Chandra}\\ 
3 & NGC 891 & [LB2005] NGC891 X1 & 02:22:31 & 42:20:30$^c$ & 11 & \citet{2005ApJS..157...59L} & {\it ROSAT}\\ 
4 & NGC 891 & [PCV2006] ULX 2 & 02:22:31.4 & 42:20:24.0$^c$ & 0.7 & This work & {\it Chandra}\\ 
5 & NGC 891 & [LB2005] NGC 891 ULX3 & 02:22:46 & 42:25:58 & 11 & \citet{2005ApJS..157...59L} & {\it ROSAT}\\ 
6 & NGC 1042 & 2XMM J024025.6-082428 & 02:40:25.6 & -08:24:29.8 & 1.4 & This work & {\it XMM-Newton}\\ 
7 & IC 342 & [LB2005] PGC 13826 ULX2 & 03:46:45.4 & 68:09:47.3 & 1.6 & This work & {\it XMM-Newton}\\ 
8 & IC 342 & CXO J034657.4+680619 & 03:46:57.4 & 68:06:19.1 & 0.9 & \citet{2010ApJS..189...37E} & {\it Chandra}\\ 
9 & NGC 2403 & RX J073655.7+653542 & 07:36:55.6 & 65:35:40.8 & 1.6 & \citet{2003AJ....125.3025S} & {\it Chandra}\\ 
10 & NGC 2500$^*$ & [SST2011] J080148.10+504354.6 & 08:01:48.1 & 50:43:54.8 & 1.1 & This work & {\it Chandra}\\ 
11 & NGC 2903 & [LB2005] NGC 2903 ULX1 & 09:32:01.9 & 21:31:11.1 & 1.6 & This work & {\it Chandra}\\ 
12 & NGC 3031$^*$ & [STS2009b] J095532.97+690033.4 & 09:55:32.9 & 69:00:33.6 & 1.8 & \citet{2011ApJS..192...10L} & {\it Chandra}\\ 
13 & NGC 3486$^*$ & XMMU J110022.4+285818 &  11:00:22.3 & 28:58:16.9 & 1.1 & \citet{2014MNRAS.442.1054H} & {\it Chandra}\\ 
14 & NGC 3521$^*$ & [SST2011] J110545.62+000016.2 &  11:05:45.6 & 00:00:16.5 & 1.6 & \citet{2011ApJS..192...10L} & {\it Chandra}\\ 
15 & NGC 3990 & 3XMM J115733.7+552711 & 11:57:33.7 & 55:27:11.1 & 1.5 & This work & {\it XMM-Newton}\\ 
16 & NGC 4190 & CXO J121345.2+363754 & 12:13:45.3 & 36:37:54.7 & 1.4 & \citet{2010ApJS..189...37E} & {\it Chandra}\\ 
17 & NGC 4258$^*$ & 3XMM J121847.6+472054 & 12:18:47.7 & 47:20:54.7 & 3.4 & This work & {\it XMM-Newton}\\ 
18 & NGC 4258 & [LB2005] NGC 4258 X9 & 12:19:23.3 & 47:09:40.4 & 1.8 & This work & {\it XMM-Newton}\\ 
19 & NGC 4490 & [SST2011] J123029.55+413927.6 & 12:30:29.5 & 41:39:27.6 & 0.9 & This work & {\it Chandra}\\ 
20 & NGC 4485 & RX J1230.5+4141 & 12:30:30.5 & 41:41:42.1 & 0.8 & This work & {\it Chandra}\\ 
21 & NGC 4490 & [SST2011] J123030.82+413911.5 & 12:30:30.7 & 41:39:11.5 & 0.9 & This work & {\it Chandra}\\ 
22 & NGC 4490 & [SST2011] J123032.27+413918.1 & 12:30:32.2 & 41:39:18.1 & 0.9 & This work & {\it Chandra}\\ 
23 & NGC 4490 & [SST2011] J123036.32+413837.8 & 12:30:36.2 & 41:38:37.9 & 0.9 & This work & {\it Chandra}\\ 
24 & NGC 4490 & CXO J123038.4+413831 & 12:30:38.2 & 41:38:31.1 & 1.3 & This work & {\it Chandra}\\ 
25 & NGC 4490 & 2XMM J123043.1+413819 & 12:30:43.2 & 41:38:18.5 & 0.9 & This work & {\it Chandra}\\ 
26 & NGC 4517 & 3XMM J123242.7+000654 & 12:32:42.7 & 00:06:54.9 & 1.4 & This work & {\it XMM-Newton}\\ 
27 & NGC 4559 & [SST2011] J123557.79+275807.4 & 12:35:57.8 & 27:58:07.4 & 1.5 & \citet{2011ApJ...741...49S} & {\it Chandra}\\ 
28 & NGC 4559 & RX J123558+27577 & 12:35:58.6 & 27:57:41.9 & 1.5 & \citet{2011ApJ...741...49S} & {\it Chandra}\\ 
29 & NGC 4594 & [LB2005] NGC 4594 X5 & 12:40:22.6 & -11:39:25.2 & 1.9 & This work & {\it XMM-Newton}\\ 
30 & NGC 4625 & [SST2011] J124152.72+411631.7 & 12:41:52.7 & 41:16:31.7 & 1.7 & This work & {\it Chandra}\\ 
31 & NGC 4631 & CXO J124157.4+323202 & 12:41:57.4 & 32:32:03.2 & 1.8 & This work & {\it Chandra}\\
32 & NGC 4631 & [LB2005] NGC 4631 ULX1 & 12:41:55 & 32:32:14$^c$ & 12 & \citet{2005ApJS..157...59L} & {\it ROSAT}\\ 
33 & NGC 4631 & [SST2011] J124155.56+323216.9 & 12:41:55.6 & 32:32:17.1$^c$ & 1.8 & This work & {\it Chandra}\\ 
34 & NGC 4631 & [WMR2006] NGC4631 XMM3 & 12:41:58.0 & 32:28:51.8 & 1.8 & This work & {\it Chandra}\\ 
35 & NGC 4631 & [SST2011] J124211.13+323235.9 & 12:42:11.1 & 32:32:36.1 & 1.8 & This work & {\it Chandra}\\ 
36 & NGC 4736 & CXO J125050.3+410712 & 12:50:50.3 & 41:07:12.2 & 1.2 & \citet{2010ApJS..189...37E} & {\it Chandra}\\ 
37 & NGC 4736 & CXO J125052.7+410719 & 12:50:52.7 & 41:07:19.0 & 1.2 & \citet{2010ApJS..189...37E} & {\it Chandra}\\ 
38 & NGC 4736 & CXO J125053.3+410714 & 12:50:53.3 & 41:07:14.0 & 1.2 & \citet{2010ApJS..189...37E} & {\it Chandra}\\ 
39 & NGC 5128 & 2XMM J132538.3-430205 & 13:25:38.3 & -43:02:04.9 & 2.3 & This work & {\it XMM-Newton}\\ 
40 & NGC 5194 & RX J132947+47096 & 13:29:47.5 & 47:09:40.7 & 1.4 & This work & {\it Chandra}\\ 
41 & NGC 5194 & [MEE95] R8 & 13:29:53.8 & 47:14:35.8 & 1.3 & This work & {\it Chandra}\\ 
42 & NGC 5457 & [LB2005] NGC 5457 X32 & 14:01:34.5 & +54:20:30.1  & 2.5 & This work & {\it XMM-Newton}\\ 
43 & NGC 5457 & CXO J140303.9+542734 & 14:03:03.9 & 54:27:33.0 & 0.8 & \citet{2010ApJS..189...37E} & {\it Chandra}\\ 
44 & NGC 5457 & CXO J140341.1+541903 & 14:03:41.2 & 54:19:03.0 & 1.0 & \citet{2010ApJS..189...37E} & {\it Chandra}\\ 
45 & NGC 5457 & [LB2005] NGC 5457 X26 & 14:04:29.2 & 54:23:53.1 & 1.9 & This work & {\it Chandra}\\ 
\hline\hline
\multicolumn{8}{l}{{\bf Notes:} $^a$:99.7$\%$ (3-$\sigma$) confidence radius around the position of the ULX within which we search for counterparts. $^b$:Catalog that provided the ULX}\\
\multicolumn{8}{l}{ position. $^c$:ULXs that have the same position within a 2-$\sigma$ confidence limit. $^*$: Sources analyzed by H14, that we re-observed in the $H$-band to provide}\\
\multicolumn{8}{l}{deeper and better seeing images. $^{\ddagger}$:Set of Identifications, Measurements and Bibliography for Astronomical Data (SIMBAD; \citealt{2000A&AS..143....9W}).}\\
\end{tabular}}
\end{center}
\end{table*}

\subsection{{\it Chandra} observations}
\label{chandra}

In an attempt to reduce the uncertainty on the X-ray position of the ULXs we 
queried the {\it Chandra} archive. For all sources with archival {\it Chandra} 
observations we used the task {\scshape acis$_{-}$process$_{-}$events} in
{\sc ciao} \citep{2006SPIE.6270E..1VF} to reprocess the event files with 
the latest calibration files (CALDB version 4.7.2) taking into account 
whether the observations were made in the "Faint" or "Very Faint" mode. We 
then produced images from data in the 0.3 -- 7 keV energy range, on which we 
run the {\sc wavdetect} task \citep{ciao} to establish accurate positions of 
all X-ray sources in the field of view of the {\it Chandra} ACIS CCDs.

In order to try to further improve the knowledge of the location of the ULXs 
we investigated whether we could apply a bore-sight correction 
(e.g. \citealt{2010MNRAS.407..645J}). For this 
we search for X-ray sources detected with {\sc wavdetect} with 
more than 20 X-ray counts and that lie within 3\arcmin\ 
of the optical axis of the satellite. 
When found, we investigated whether counterparts in the Sloan Digital Sky Survey 
(SDSS, \citealt{2003AJ....125.1559P}) or 
2MASS catalogue exist. We considered the SDSS or 2MASS source a counterpart
to the X-ray source if the offset between the catalogue coordinates and the X-ray
source position is less than 1\arcsec\ . We then apply shifts in R.A.~and Dec.~to the 
X-ray coordinates using the {\sc wcs$_{-}$update} tool.
Using this bore-sight correction procedure we were able to reduce the 
uncertainty on the location of 15 ULXs.
Crowding of the optical or NIR fields precluded us to match the 
X-ray source to a unique optical/NIR source in 6 cases.

For 4 ULXs only {\it ROSAT} coordinates are available in the literature 
(ULXs with IDs 11, 40, 41, and 45 in Table~\ref{tab:ulxs}). For these sources 
we provide the {\it Chandra} positions in Table~\ref{tab:corrected}. Even though 
we could not apply a bore-sight correction, those are more accurate than 
the {\it ROSAT} source positions from \citet{2005ApJS..157...59L}.

To evaluate the uncertainties in the final position of the ULX, we
have to take into account: (a) the statistical uncertainty given by 
{\scshape wavdetect} in the source localization on the CCD, (b) the WCS uncertainty
of the NIR image (see Table~\ref{tab:galaxies}). In addition, for those sources where we 
applied a bore-sight correction, we must take into account
(c) the statistical uncertainty 
between the corrected {\scshape wavdetect} position of the 
optical/infrared counterpart with its 
position in the SDSS/2MASS catalogs, (d) and the instrinsic 
uncertainties of the SDSS/2MASS catalogs with respect
to the ICRS: 3 mas (systematic) and 6 mas (statistical) for SDSS 
\citep{2003AJ....125.1559P} and 15 mas for 2MASS. We added systematic 
uncertainties linearly and statistical uncertainties quadratically. 
The 99.7$\%$ (3-$\sigma$) confidence radii for the ULXs final position
are indicated in Table~\ref{tab:ulxs}.

\subsection{{\it XMM-Newton} observations}
\label{xmm}

Nine ULXs (ID 6, 7, 15, 17, 18, 26, 29, 39 and 42) were observed by {\it XMM-Newton}
and therefore, analyzed with the Science Analysis System 
(SAS, \href{http://www.cosmos.esa.int/web/xmm-newton/sas}{{\it XMM-Newton} Data
Analysis}), using the {\scshape catcorr} task. The SAS software
updates the position of the sources by cross-matching the positions
with three catalogs: SDSS, 2MASS and the US Naval Observatory (USNO-B1.0,
\citealt{2003AJ....125..984M})
to find optical or infrared counterparts and applying shifts or 
rotations to optimize the match.

The uncertainties of the final positions are derived with the systematic
error given by {\scshape catcorr}, which takes into account the uncertainty
of the corrected position and the rotation of the field of view; 
the WCS uncertainty of the NIR image (from Table~\ref{tab:galaxies}); and the 
intrinsic uncertainties of the SDSS/2MASS/USNO-B1.0 catalogs
with respect to the ICRS: 0.03\arcsec\ and 0.06\arcsec\ for SDSS, 0.015\arcsec\ for 
2MASS and 0.2\arcsec\ (systematic) for USNO-B1.0 \citep{2003AJ....125..984M}.
The 99.7$\%$ (3-$\sigma$) confidence radii for the ULXs final position
are given in Table~\ref{tab:ulxs}.

\subsection{{\it ROSAT} observations}
\label{rosat}

There are 3 ULXs (ID 3, 5, and 32) which have been
observed by the {\it ROSAT} mission but not by {\it Chandra} or {\it XMM-Newton}. 
Because of this, we could not
improve the astrometry on these sources positions beyond that available
in the literature. The 1-$\sigma$
uncertainty on these positions is reported to be 3.62\arcsec\ 
by \citet{2005ApJS..157...59L}.
The 99.7$\%$ confidence radii for the these positions are provided 
in Table~\ref{tab:ulxs}, and are calculated taking also into account the WCS
uncertainty of the NIR image.

\begin{figure}
    \begin{minipage}{0.25\textwidth}
        \includegraphics[width=\textwidth]{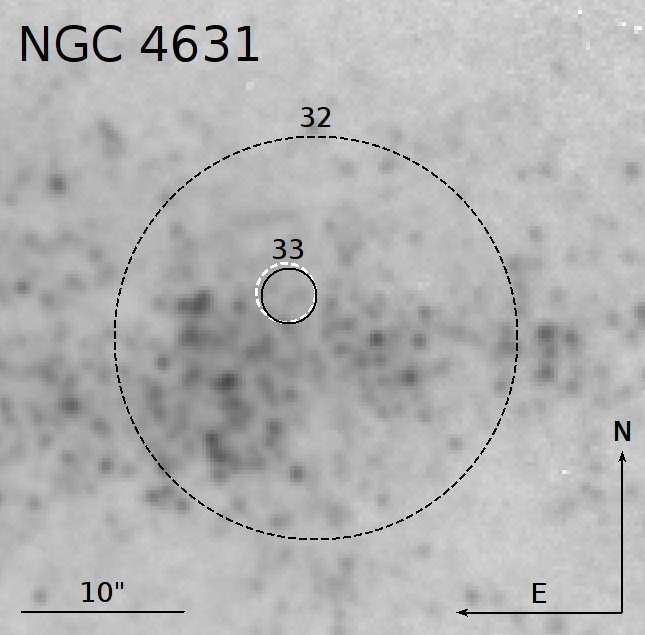}
    \end{minipage}%
    \begin{minipage}{0.25\textwidth}
        \includegraphics[width=0.99\textwidth]{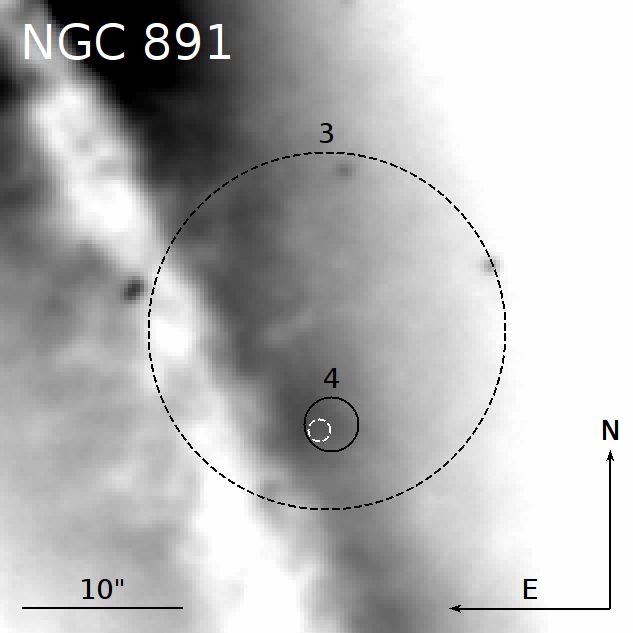}
    \end{minipage}%
\caption{{\it ROSAT} (black dashed circle) and {\it Chandra} (black solid circle) positions of two ULXs on the NIR $H$-band image. The radii of the circles indicate the 99.7$\%$ confidence level on the location of the ULX. The position of these ULXs coincide to a 95$\%$ (2-$\sigma$) confidence level; thus, we deem it likely that the {\it ROSAT} sources are the same as the {\it Chandra} sources and we remove the 2 {\it ROSAT} sources from our list. The white dashed circle indicate the position obtained after performing a boresight correction on the {\it Chandra} positions (see section~\ref{chandra}).}
    \label{fig:sameulx}
\end{figure}

The position of two of the ULXs observed by {\it ROSAT} 
(NGC 4631 ID 32 and NGC 891 ID 3, from \citealt{2005ApJS..157...59L})
coincide with the position of two unique ULXs observed by {\it Chandra} 
(NGC 4631 ID 33 and NGC 891 ID 4, from \citealt{2011ApJ...741...49S})
to a 95$\%$ (2-$\sigma$) confidence level (see Figure~\ref{fig:sameulx}).
For these two ULXs observed by {\it Chandra} we performed a boresight correction (see
section~\ref{chandra}), obtaining a 99.7$\%$ confidence radii of
0.93\arcsec\ and 0.71\arcsec\, for NGC 4631 ID 33 and NGC 891 ID 4 respectively. 
Hence, it is likely that NGC 4631 ID 32 and ID 33 
are the same source, as well as NGC 891 ID 3 and ID 4. For the detection of candidate counterparts
for these two ULXs, we will use the {\it Chandra} localization.

\section{Results and discussion}
\label{results}

Of the 43 ULXs we observed, 15 have one NIR candidate counterpart,
and 1 has two candidate counterparts (see Figure~\ref{fig:images}).
Their apparent and absolute $H$-band magnitudes are detailed in 
Table~\ref{tab:candidates}.
For the cases where no counterpart was detected, the limiting (absolute)
magnitude of the NIR image is indicated.

For the ULXs with candidate counterparts, we can investigate the nature of
these by means of their absolute magnitude. Eight of the candidate counterparts have
magnitudes ranging from -8.66 mag to -11.18 mag, and within the uncertainties,
these magnitudes are consistent with those of a RSG. 
The remaining 9 candidate counterparts have absolute magnitudes in the range
-11.44 mag to -14.85 mag, and are therefore, too bright to be a single RSG. 
These sources are most likely background AGN or unresolved star clusters. 
However, we need to 
search for further evidence to determine their 
true nature. We want to stress that the only way to be certain of the classification
of each candidate is to take spectra, but below we provide some clues on their 
possible nature.

For the 27 ULXs without a detected counterpart, the apparent limiting
magnitudes range from 18.09 to 20.35 mag in the $H$-band. The limiting 
magnitudes brighter than 20 mag are mainly caused by crowded fields or high
background from the host galaxies (i.e. NGC 4490, NGC 891, NGC 4736) and it
will be difficult to improve on these using ground based, natural seeing limited
observations.

\subsection{Re-observed ULXs from H14}
\label{marianne}

We re-observed 6 ULXs previously studied in H14: NGC 855 ULX ID 1,
NGC 2500 ULX ID 10, NGC 3031 ULX ID 12, NGC 3486 ULX ID 13, NGC 3521 ULX
ID 14 and NGC 4258 ULX ID 17. 
We performed all our observations in the $H$-band, while H14 took
$K_s$ band images, with the exception of NGC 4258, which they also observed in the $H$-band. 
Our conclusions do not differ from H14 on ULX IDs 1, 10, 12 and 13.
For the two other sources we provide a more detailed assessment below.

\subsubsection{NGC 3521 ULX ID 14}
\label{ulx14}

The 99.7$\%$ confidence radius given by H14
for NGC 3521 is 2.1\arcsec\ , whereas our value is 1.6\arcsec\ (see Figure~\ref{fig:3521ulx14}). 
We detect a candidate counterpart 
(absolute $H$-band magnitude of -10.93 $\pm$ 0.93 mag), in contrast to 
the reported $K_s$ non-detection by H14. Our candidate has an absolute magnitude
consistent with an M-type supergiant. The limiting magnitude for our NIR image is 
20.17 $\pm$ 0.18 mag, and theirs is 19.25 mag; our average seeing is 0.9\arcsec\ and theirs is 0.7\arcsec\ . Due to the approximate $H$-$K$ = 0 color or RSGs, our detection is consistent with their non-detection.

\subsubsection{NGC 4258 ULX ID 17}
\label{ulx17re}

Our 99.7$\%$ confidence radius around the ULX position is 3.4\arcsec\ , whereas the radius quoted
in H14 is 2\arcsec\ . For this ULX neither we nor H14 found enough astrometric standard stars 
(see subsection~\ref{datared}) local to the ULX, so we both used the 
global astrometric solution that {\scshape theli} provides. 
The uncertainty on the position of the ULX 
is larger than that of H14, thus, our 99.7$\%$ confidence radius is larger.
The average seeing in our NIR image is 0.8\arcsec\ , in contrast with the 0.7\arcsec\ 
seeing from H14. However, our limiting magnitude is almost
1 mag deeper than the one found by H14 and we find a candidate counterpart, 
whereas they indicated
a non-detection. The detected candidate counterpart
(see Figure~\ref{fig:re4258}) is located 2\arcsec\  from the center of the ULX
 position, and has an absolute $H$-band magnitude of -9.26 $\pm$ 0.18 mag, making it a 
 potential RSG candidate. This $H$-band source lies outside the error radius
 from H14, but inside our 99.7$\%$ confidence radius, hence, we deem it a candidate counterpart.
 This ULX lies in the outskirts of the spiral arms of NGC 4258.

\subsection{NIR non-detections for ULXs with published searches for optical or radio counterparts}
\label{optical/radio}

Eight of the ULXs in our sample appear in searches for optical or radio candidate counterparts.
Of these, we detect NIR candidate counterparts for 3 ULXs, which we discuss in detail in 
subsections~\ref{RSG} and~\ref{AGN}. We do not detect any NIR candidate counterpart for the
other 5 ULXs: IC 342 ULX ID 8, NGC 3031 ULX ID 12, NGC 4190 ULX ID 16, NGC 4490 ULX ID 21 and 
NGC 4736 ULX ID 36.

IC 342 ULX ID 8 was observed by \citet{2013ApJS..206...14G}, where they detected one
optical candidate counterpart. NGC 3031 ULX ID 12 was observed and analyzed by 
\citet{2002ApJ...580L..31L} and \citet{2013ApJS..206...14G}, 
and both reported an optical candidate counterpart. 
NGC 4190 ULX ID 16 was observed also by \citet{2013ApJS..206...14G}, 
and they detected an optical candidate counterpart. \citet{2006IAUS..230..310G} observed 
NGC 4490 ULX ID 21 and detected a candidate counterpart HII region. In contrast, we do not detect
any NIR candidate counterpart for these 3 ULXs. To investigate what our 
non-detections imply for the nature of the optical candidate counterparts,  we need to be able
to assess the probability that these optical and NIR sources are associated and for that one needs
the coordinates of the optical candidate counterparts. Unfortunately, they are no given in the 
literature for these sources.

NGC 4736 ULX ID 36 was analyzed by \citet{2013ApJS..206...14G}, and was found
to not have an optical candidate counterpart. We also find no NIR candidate counterpart.

\begin{table*}
\vspace{5mm}
\begin{center}
\caption{NIR candidate counterparts to the ULXs listed in Table~\ref{tab:ulxs} and our preliminary classification as candidate RSG, AGN, or stellar cluster (SC). The preliminary classification is based on their absolute magnitude, WISE colours, visual inspection of the NIR image and/or spatial extent of the candidate counterpart.}
\label{tab:candidates}
\resizebox{\textwidth}{!}{\begin{tabular}{|lcccccccc|}
\hline\hline
Galaxy & ULX ID & Detection & R.A & Dec. & Position$^a$ & Apparent & Absolute$^b$ & Preliminary\\
 & & & & & uncertainty & magnitude & magnitude & classification\\
 & & (y/n) & (hh:mm:ss) & (dd:mm:ss) & ( \arcsec\ ) & (mag) & (mag) &\\ 
\hline\hline
NGC 855 & 1 & n &-&-&-& $>$ 18.37 $\pm$ 0.12 & $>$ -11.48 $\pm$ 0.51 & -\\
NGC 891 & 2 & n &-&-&-& $>$ 19.11 $\pm$ 0.51 & $>$ -11.20 $\pm$ 0.26 &-\\
NGC 891 & 4 & n &-&-&-& $>$ 19.11 $\pm$ 0.51 & $>$ -11.20 $\pm$ 0.26 &-\\
NGC 891 & 5 & n &-&-&-& $>$ 20.06 $\pm$ 0.36 & $>$ -10.10 $\pm$ 0.20 &-\\
NGC 1042 & 6 & n &-&-&-&  $>$ 19.55 $\pm$ 0.35 & $>$ -8.22 $\pm$ 0.13 &-\\
IC 342 & 7 & y & 03:46:45.16 & +68:09:48.73 & 0.38 & 18.52 $\pm$ 0.05 & -8.66 $\pm$ 0.12 & RSG\\
IC 342 & 8 & n &-&-&-& $>$ 19.31 $\pm$ 0.41 & $>$ -8.08 $\pm$ 0.13 &-\\
NGC 2403 & 9 & y & 07:36:55.39 & +65:35:41.72 & 0.81 & 17.46 $\pm$ 0.02 & -10.05 $\pm$ 0.26 & RSG\\
NGC 2500 & 10 & n &-&-&-& $>$ 20.16 $\pm$ 0.06 & $>$ -9.52 $\pm$ 0.19 &-\\
NGC 2903 & 11 & n &-&-&-& $>$ 20.17 $\pm$ 0.07 & $>$ -9.58 $\pm$ 0.75 &-\\
NGC 3031 & 12 & n &-&-&-& $>$ 18.94 $\pm$ 0.04 & $>$ -8.69 $\pm$ 0.11 &-\\
NGC 3486 & 13 & n &-&-&-& $>$ 19.03 $\pm$ 0.27 & $>$ -10.05 $\pm$ 0.20 &-\\
NGC 3521 & 14 & y & 11:05:45.62 & +00:00:17.70 & 0.38 & 19.83 $\pm$ 0.06 & -10.93 $\pm$ 0.93 & RSG\\
NGC 3990 & 15 & n &-&-&-& $>$ 19.11 $\pm$ 0.19 & $>$ -10.71 $\pm$ 0.47 &-\\
NGC 4190 & 16 & n &-&-&-& $>$ 19.06 $\pm$ 0.01 & $>$ -8.01 $\pm$ 0.31 &-\\
NGC 4258 & 17 & y & 12:18:47.73 & +47:20:53.08 & 2.55 & 20.06 $\pm$ 0.07 & -9.26 $\pm$ 0.18 & RSG\\
NGC 4258 & 18 & y & 12:19:22.98 & +47:09:31.35 & 0.85 & 15.86 $\pm$ 0.001 & -13.46 $\pm$ 0.23 & AGN\\
NGC 4490 & 19 & y & 12:30:29.54 & +41:39:26.71 & 0.42 & 16.05 $\pm$ 0.01 & -13.41 $\pm$ 0.27 & SC\\
NGC 4485 & 20 & y & 12:30:30.45 & +41:41:42.79 & 0.09 & 19.01 $\pm$ 0.06 & -10.69 $\pm$ 0.41 & RSG\\
NGC 4490 & 21 & n &-&-&-& $>$ 18.48 $\pm$ 0.62 & $>$ -10.36 $\pm$ 0.21 &-\\
NGC 4490 & 22 & n &-&-&-& $>$ 18.48 $\pm$ 0.62 & $>$ -10.36 $\pm$ 0.21 &-\\
NGC 4490 & 23 & n &-&-&-& $>$ 18.48 $\pm$ 0.62 & $>$ -10.36 $\pm$ 0.21 &-\\
NGC 4490 & 24 & y & 12:30:38.24 & +41:38:31.59 & 0.42 & 16.69 $\pm$ 0.008 & -12.77 $\pm$ 0.27 & 2 RSGs\\
NGC 4490 & 25 & y & 12:30:43.06 & +41:38:18.98 & 0.42 & 18.41 $\pm$ 0.04 & -11.04 $\pm$ 0.27 & SC\\
NGC 4517 & 26 & n &-&-&-& $>$ 19.19 $\pm$ 0.51 & $>$ -9.22 $\pm$ 0.19 &-\\
NGC 4559 & 27$^c$ & y & 12:35:57.72 & +27:58:07.71 & 0.67 & 17.33 $\pm$ 0.01 & -11.99 $\pm$ 0.90 & SC\\
NGC 4559 & 27$^c$ & y & 12:35:57.84 & +27:58:06.93 & 0.67 & 17.29 $\pm$ 0.01 & -12.03 $\pm$ 0.90 & SC\\
NGC 4559 & 28 & y & 12:35:58.65 & +27:57:41.45 & 0.67 & 18.95 $\pm$ 0.05 & -10.37 $\pm$ 0.89 & RSG\\
NGC 4594 & 29 & y & 12:40:22.70 & -11:39:24.07 & 0.56 & 18.65 $\pm$ 0.02 & -11.61 $\pm$ 0.57 & AGN\\
NGC 4625 & 30 & n &-&-&-& $>$ 19.65 $\pm$ 0.05 & $>$ -9.67 $\pm$ 0.18 &-\\
NGC 4631 & 31 & n &-&-&-& $>$ 18.53 $\pm$ 0.27 & $>$ -10.03 $\pm$ 0.34 &-\\
NGC 4631 & 33 & n &-&-&-& $>$ 18.53 $\pm$ 0.27 & $>$ -10.03 $\pm$ 0.34 &-\\
NGC 4631 & 34 & n &-&-&-& $>$ 18.53 $\pm$ 0.27 & $>$ -10.03 $\pm$ 0.34 &-\\
NGC 4631 & 35 & y & 12:42:11.11 & +32:32:37.12 & 1.54 & 14.48 $\pm$ 0.001 & -14.85 $\pm$ 0.49 & SC\\
NGC 4736 & 36 & n &-&-&-& $>$ 18.31 $\pm$ 0.04 & $>$ -9.76 $\pm$ 0.06 &-\\
NGC 4736 & 37 & n &-&-&-& $>$ 18.31 $\pm$ 0.04 & $>$ -9.76 $\pm$ 0.06 &-\\
NGC 4736 & 38 & n &-&-&-& $>$ 18.31 $\pm$ 0.04 & $>$ -9.76 $\pm$ 0.06 &-\\
NGC 5128 & 39 & n &-&-&-& $>$ 17.79 $\pm$ 0.11 & $>$ -9.72 $\pm$ 0.12 &-\\
NGC 5194 & 40 & y & 13:29:47.50 & +47:09:40.78 & 0.55 & 18.60 $\pm$ 0.04 & -11.18 $\pm$ 0.04 & RSG\\
NGC 5194 & 41 & n &-&-&-& $>$ 18.09 $\pm$ 0.01 & $>$ -11.38 $\pm$ 0.01 &-\\
NGC 5457 & 42 & y & 14:01:32.98 & +54:20:41.47 & 0.18 & 18.67 $\pm$ 0.02 & -10.54 $\pm$ 0.13 & AGN\\
NGC 5457 & 43 & n &-&-&-& $>$ 20.35 $\pm$ 0.10 & $>$ -8.91 $\pm$ 0.14 &-\\
NGC 5457 & 44 & n &-&-&-& $>$ 20.35 $\pm$ 0.10 & $>$ -8.91 $\pm$ 0.14 &-\\
NGC 5457 & 45 & y & 14:04:29.20 & +54:23:51.83 & 1.01 & 17.78 $\pm$ 0.01 & -11.44 $\pm$ 0.22 & SC\\
\hline\hline
\multicolumn{9}{l}{{\bf Notes:} $^a$:99.7$\%$ (3-$\sigma$) confidence radius around the position of the NIR candidate counterpart; this value is calculated taking into account the}\\
\multicolumn{9}{l}{uncertainties of the astrometric correction on the NIR images and the uncertainties given by the source detection by {\scshape SExtractor}. We added}\\
\multicolumn{9}{l}{systematic uncertainties linearly and statistical uncertainties quadratically. $^b$:Values calculated using the distance from Table~\ref{tab:galaxies} and the apparent}\\
\multicolumn{9}{l}{magnitude, with a 1-$\sigma$ uncertainty. $^c$:Multiple NIR candidate counterparts detected.}\\
\end{tabular}}
\end{center}
\end{table*}

\subsection{NIR Red supergiant candidates}
\label{RSG}

\subsubsection{IC 342 ULX ID 7}
\label{ulx7}

Just outside the 99.7$\%$ confidence radius
there is a NIR candidate counterpart detected
(see Figure~\ref{fig:manycandidates}), with an absolute magnitude of -8.66 $\pm$ 0.12,
consistent with the absolute magnitude of RSGs \citep{1985ApJS...57...91E,2000asqu.book..381D}.
Visual inspection suggests that this source has several candidate counterparts inside the confidence radius; 
however, {\scshape SExtractor} detected only a single object in this crowded area.

This source was detected in {\it ROSAT} HRI observations by \citet{2005ApJS..157...59L}, who both interpolated the {\it ROSAT} flux into a wider 0.3--8 keV band and corrected it for Galactic foreground column (based on a power-law spectrum with $\Gamma = 1.7$, and no additional absorption) to obtain a luminosity of $1.5 \times 10^{39}$ erg/s for the object (based on a distance of 3.9 Mpc).  Subsequent catalogued {\it XMM-Newton} detections of this source have reported fluxes consistent with much lower luminosities, ranging between $\sim 10^{37}$ \citep{2003MNRAS.346..265K} and $\sim 5 \times 10^{38}$ erg/s \citep{2012ApJ...756...27L} for an improved distance of 2.73 Mpc.
\citet{2012ApJ...756...27L} classified it as a 
Super Soft X-ray Source, based on its hardness ratio. Additionally, \citet{2003AJ....126.2797B} determined that this X-ray source is probably not intrinsic to IC 342, but a faint foreground star,  with a position that falls just 0.3\arcsec\ from the best fitted X-ray position. These newly 
determined lower luminosities (if in IC 342) and the possible identification as a foreground star make that we do not consider this source a ULX.

\subsubsection{NGC 2403 ULX ID 9}
\label{ulx9}

ULX ID 9 lies in NGC 2403 (see Figure~\ref{fig:2403ulx9}), a bulge-less galaxy.
We detect a single NIR candidate counterpart of -10.05 $\pm$ 0.26 mag, consistent with 
a RSG absolute magnitude. Even though {\scshape SExtractor} detected it as a single 
object in this crowded area, visual inspection of the image suggests that there 
is a second fainter NIR source.

\subsubsection{NGC 4485 ULX ID 20}
\label{ulx20}

For NGC 4485 ULX ID 20 we find a -10.69 $\pm$ 0.41 mag candidate counterpart 
(see Figure~\ref{fig:4485ulx20}), which is consistent
with a RSG. This galaxy is interacting with NGC 4490, and the galaxies have been
shown to have a high star formation rate \citep{2002MNRAS.337..677R}.

\subsubsection{NGC 4490 ULX ID 24}
\label{ulx24}

NGC 4490 ULX ID 24 is located 24\arcsec\ (810 pc) from the galactic nucleus, 
and its -12.77 $\pm$ 0.27 mag candidate 
counterpart (see Figure~\ref{fig:4490ulx24}) seems to be an extended object, 
not a point source. 
Visual inspection suggests that this object consists of 2 unresolved objects, 
possibly 2 RSGs.
{\scshape SExtractor} detected it as a single object in this crowded area.

\subsubsection{NGC 4559 ULX ID 28}
\label{ulx28}

We find one NIR candidate counterpart for NGC 4559 ULX ID 28 
(see Figure~\ref{fig:4559ulx28}), with an absolute magnitude of
-10.37 $\pm$ 0.89 mag, consistent with a RSG. 
This ULX was observed in the optical by different authors.
\citet{2004MNRAS.349...39C} did not detect any candidate counterparts for this ULX.
\citet{2006ApJS..166..154P} identified two
candidate counterparts with $V$-band apparent magnitudes 24.4 and 23.7, and
located 0.65\arcsec\  and 1.09\arcsec\  from their best-fit X-ray position, respectively.
Given that these are
smaller than our 99.7$\%$ confidence radius (1.5\arcsec\ ), both sources would fall in 
our error region. However, without the coordinates of these
optical candidate counterparts, we cannot be sure if our 16.69 $\pm$ 0.008 
NIR candidate counterpart is associated to any of them.

Additionally, \citet{2016arXiv160603024V} identified one optical candidate counterpart 
with $V$-band apparent magnitude of 24.04, located less than 0.15\arcsec\ from 
the ULX, and an estimated reddening of E(B-V) = 0.26 $\pm$ 0.06. Comparing their optical and our NIR source positions, we find that they are 2.314\arcsec\ apart, making it likely that one of the several optical sources present in the 2\arcsec\ box in Figure 1 of \citet{2016arXiv160603024V} is a RSG that we detect in the NIR. Since their error circle for the X-ray source is smaller than ours, it would exclude our RSG as a candidate counterpart.

\subsubsection{NGC 5194 ULX ID 40}
\label{ulx40}

ULX ID 40 is located in NGC 5194, approximately 3\arcsec\ (110 pc in projection) from the 
star cluster [HL2008] 21287 \citep{2008AJ....135.1567H}. 
We find a NIR candidate counterpart (see Figure~\ref{fig:5194ulx40})
with absolute magnitude -11.18 $\pm$ 0.04 mag, consistent with a RSG absolute 
magnitude within the 1-$\sigma$ confidence limit.

\subsection{NIR Background AGN candidates}
\label{AGN}

\subsubsection{NGC 4258 ULX ID 18}
\label{ulx18}

This ULX has one candidate counterpart 
(see Figure~\ref{fig:4candidates}) within the 99.7$\%$ confidence radius, with
absolute magnitude of -13.46 $\pm$ 0.23 mag, which has been observed with the 
Wide-Field Infrared Survey Explorer (WISE, \citealt{2010AJ....140.1868W}).
Even though the WISE best angular resolution is 6.1\arcsec\, 
the association and photometry are robust since the field of view of the ULX is not crowded.
The WISE colours of this candidate are [3.4]-[4.6] = 0.78 and 
[4.6]-[12] = 2.69, making it an AGN according to the
WISE colour-colour diagrams from \citet{2012ApJ...748...68D} and \citet{2015ApJ...798...38S}, 
which makes it
likely that the X-ray source classified as a ULX is associated with the AGN.

\subsubsection{NGC 4594 ULX ID 29}
\label{ulx29agn}

For NGC 4594 ULX ID 29 (see Figure~\ref{fig:4594ulx29}) the field of view is also not
crowded, so we can rely on the WISE data to further investigate the nature of the
detected candidate counterpart. This NIR source (-11.61 $\pm$ 0.57 mag)
has colours in the WISE bands of [3.4]-[4.6] = 1.39 and [4.6]-[12] = 2.72,
classifying it as a Seyfert galaxy according to the diagrams of \citet{2012ApJ...748...68D} 
and \citet{2015ApJ...798...38S}.
Therefore, we deem it likely that this ULX is in reality a background AGN.

\subsubsection{NGC 5457 ULX ID 42}
\label{ulx42}

We find one NIR candidate counterpart for ULX ID 42 (see Figure~\ref{fig:5457ulx42}), 
with an absolute magnitude of -10.54 $\pm$ 0.13 mag. 
Visual inspection of the NIR candidate counterpart shows an apparently extended object
that is located nowhere near a star forming region. Therefore we deem it likely
that the detected source is the host galaxy of a background AGN 
(which would explain its absolute magnitude). 

\subsection{NIR Star cluster candidates}
\label{star}

\subsubsection{NGC 4490 ULX ID 19}
\label{ulx19}

\citet{2011AN....332..384P} analyzed NGC 4490 ULX 19 and found 1 radio candidate counterpart
for it, which was originally observed with {\it Chandra}
(source CXOU J123029.5+413927, \citealt{2002MNRAS.337..677R}).
We detect one NIR candidate counterpart inside the 99.7$\%$ confidence radius around the ULX
position (see Figure~\ref{fig:4490ulx19}) with an absolute $H$-band magnitude of 
-13.41 $\pm$ 0.27 mag.  The absolute magnitude and the spatial extent of this NIR object 
suggests that it is a star cluster.

\subsubsection{NGC 4490 ULX ID 25}
\label{ulx25}

For NGC 4490 ULX ID 25 (see Figure~\ref{fig:4490ulx25}) we detect one NIR candidate counterpart
(-11.04 $\pm$ 0.27 mag) which appears to be an extended source. 
It is possible that the NIR candidate 
counterpart is a star cluster, i.e. a group of unresolved RSGs.
This ULX was observed before by \citet{2006ApJS..166..154P}, where
they reported 6 optical candidate counterparts. 
However, all these optical candidate sources fall outside the 99.7$\%$ confidence error 
radius for this ULX, and thus, none is associated with the NIR source reported here.

\subsubsection{NGC 4559 ULX ID 27}
\label{ulx27}

We detect 2 NIR candidate counterparts for NGC 4559 ULX ID 27 
(see Figure~\ref{fig:4559ulx27}), with absolute magnitudes
of -11.99 $\pm$ 0.90 mag and -12.03 $\pm$ 0.90 mag. 
Even though the area is not crowded, we cannot use WISE data 
to investigate the nature of these two candidate counterparts as they are only
1.5\arcsec\  apart. Therefore, unless the colours of both candidate counterparts are such that only one source
determines the WISE photometry, the WISE data will be a blend of the two sources.
Visual inspection of the light distribution of the NIR image, taking into account the absolute magnitudes
of the NIR sources, suggests to us that these candidate counterparts are probably star clusters.

\subsubsection{NGC 4631 ULX ID 35}
\label{ulx35}

ULX ID 35 lies in a crowded area 
(see Figure~\ref{fig:4631ulx35}). The NIR candidate counterpart
is extremely bright (-14.85 $\pm$ 0.49 mag), so we rule it out as a single RSG. 
Since it appears to be an extended object, it could be a star cluster.

\subsubsection{NGC 5457 ULX ID 45}
\label{ulx45}

We detected a NIR candidate counterpart for NGC 5457 ULX ID 45 
(see Figure~\ref{fig:5457ulx45}), with absolute magnitude
of -11.44 $\pm$ 0.22 mag. The candidate seems to be an extended object, so it could be an 
unresolved young star cluster.

\section{Conclusions}
\label{conclusions}

This is the second paper on our systematic search for NIR counterparts of ULXs.
We observed 42 ULXs in the $H$-band and detected candidate counterparts for
15 of them. Of these, one has 2 multiple candidate counterparts.
For the ULXs with non-detections, we report their limiting magnitudes.

We find that 7 ULXs have NIR candidate counterparts with absolute magnitudes consistent
with that of a single RSG. Two of these (NGC 3521 ULX ID 14 and NGC 4258
ULX ID 17) correspond to ULXs observed
before by H14, and that we re-observed in 
this work. H14 reported these ULXs as having no candidate counterparts.
We find candidate counterparts because our limiting magnitudes for those two
NIR images are deeper than the ones reported by H14.
Another one of our 7 ULXs with NIR RSG candidates correspond to NGC 4559 ULX 28,
which has known optical candidate counterparts \citep{2006ApJS..166..154P,2016arXiv160603024V}.
To be completely certain of the nature of these NIR RSG candidate
counterparts, spectroscopic confirmation is required.

We find 8 ULXs with NIR candidate counterparts with absolute magnitudes too bright to 
be single stars. Indeed, some of them are extended sources. For three of these ULXs
we deem it likely that they are the host galaxies of 
background AGNs, based on WISE data of the field and the 
colour-colour diagrams of \citet{2012ApJ...748...68D} and \citet{2015ApJ...798...38S}, 
and their spatial extent.
After visual inspection of the NIR images, we conclude that the remaining 5 ULXs 
are more likely to be star clusters. 

We remove three sources from our list (see Table~\ref{tab:ulxs}). 
Our more accurate positions for NGC 891 ULX ID 3 and 4, NGC 4631 ULX ID 32 and 33, show that
their positions are consistent with them being the same at the 2-$\sigma$ level. Thus, we
conclude that the {\it ROSAT} sources are the same as the {\it Chandra}
sources. The third removed source is IC 342 ULX ID 7, originally classified as a ULX 
by \citet{2005ApJS..157...59L}, but a new distance estimate to IC 342 renders the 
luminosity to fall below the ULX limit. Furthermore, it might be associated with 
a foreground star \citep{2003AJ....126.2797B}.

We detect counterparts for 36$\%$ of our total sample, where 17$\%$
corresponds to RSG candidates and 19$\%$ to AGN/star clusters. These
values are similar to the ones from H14, where
27$\%$ of their ULXs present counterparts, and 18$\%$ are RSG candidates and 11$\%$
corresponds to AGN/star clusters.

\section*{Acknowledgements}
This research is based on observations made with the William Herschel Telescope operated on the island of La Palma by the Isaac Newton Group in the Spanish Observatorio del Roque de los Muchachos of the Instituto de Astrof\'{i}sica de Canarias.
We have made use of the SIMBAD database, operated at CDS, Strasbourg, France; of the NASA/IPAC Extragalactic Database (NED) which is operated by the Jet Propulsion Laboratory, California Institute of Technology, under contract with the National Aeronautics and Space Administration; and of data obtained from the Chandra Data Archive and the Chandra Source Catalog, and software provided by the Chandra X-ray Center (CXC) in the application packages CIAO, ChIPS, and Sherpa. PGJ and KML acknowledge funding from the European Research Council under ERC Consolidator Grant agreement no 647208. TPR acknowledges funding from STFC as part of the consolidated grant ST/L00075X/1.

\begin{figure*}
    \begin{minipage}{0.333\textwidth}
        \includegraphics[width=\textwidth]{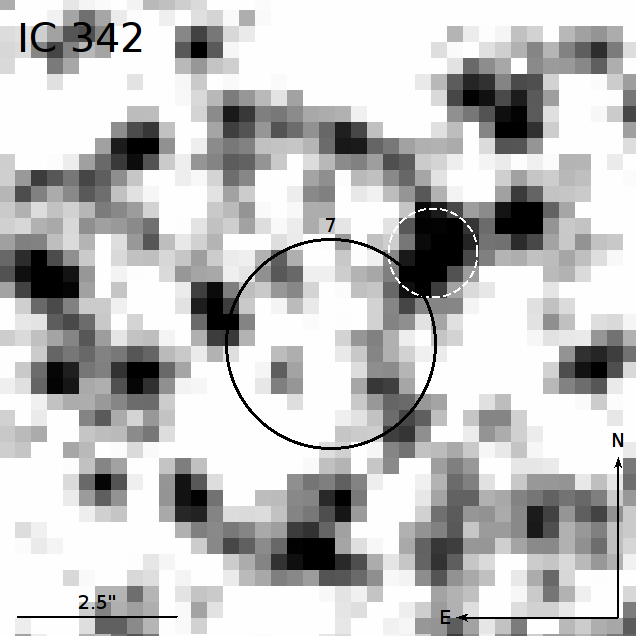}%
        \subcaption{IC 342, ULX ID 7, 3-$\sigma$ = 1.6"}
        \label{fig:manycandidates}
         \vspace{0.5cm}
    \end{minipage}%
    \begin{minipage}{0.333\textwidth}
        \includegraphics[width=\textwidth]{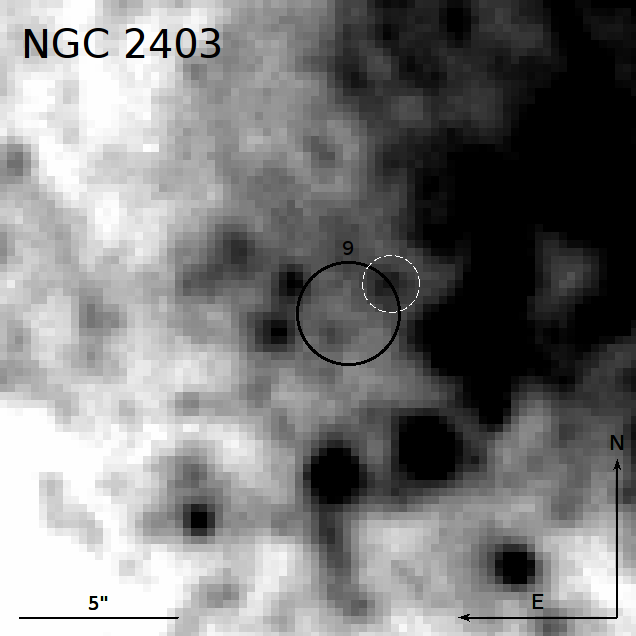}%
        \subcaption{NGC 2403, ULX ID 9, 3-$\sigma$ = 1.6"} 
        \label{fig:2403ulx9}
        \vspace{0.5cm}       
    \end{minipage}%
    \begin{minipage}{0.333\textwidth}
        \includegraphics[width=\textwidth]{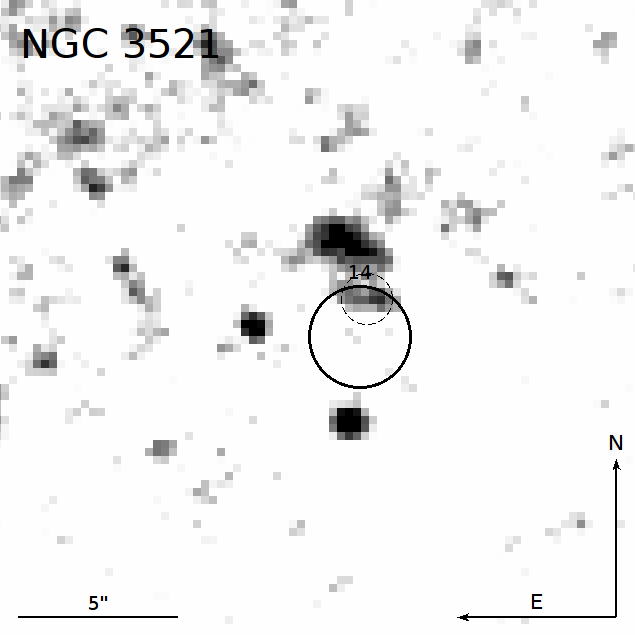}%
        \subcaption{NGC 3521, ULX ID 14, 3-$\sigma$ = 1.6"} 
        \label{fig:3521ulx14}
         \vspace{0.5cm}               
    \end{minipage}
    \begin{minipage}{0.333\textwidth}
        \includegraphics[width=\textwidth]{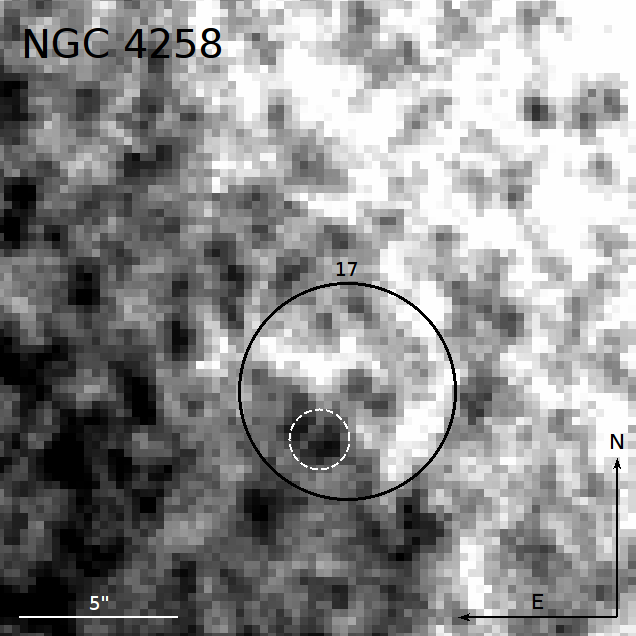}%
         \subcaption{NGC 4258, ULX ID 17, 3-$\sigma$ = 3.4"} 
         \vspace{0.5cm}
         \label{fig:re4258}               
    \end{minipage}%
    \begin{minipage}{0.333\textwidth}
        \includegraphics[width=\textwidth]{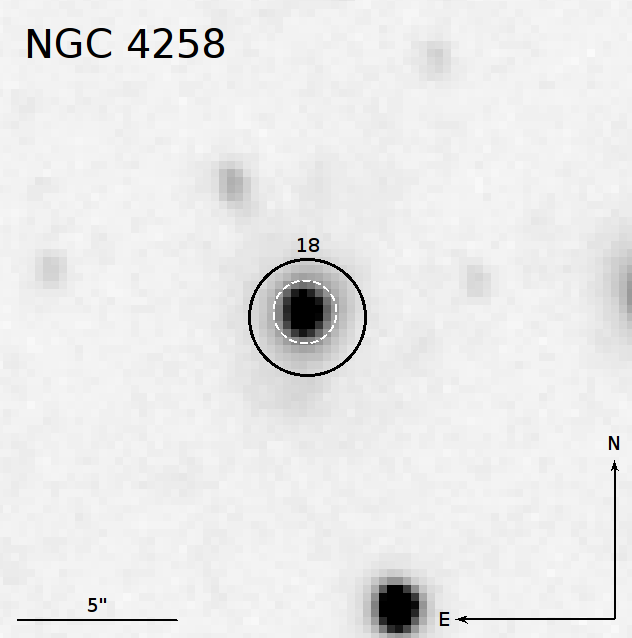}%
        \subcaption{NGC 4258, ULX ID 18, 3-$\sigma$ = 1.8"}
         \vspace{0.5cm} 
         \label{fig:4candidates}               
    \end{minipage}%
     \begin{minipage}{0.333\textwidth}
        \includegraphics[width=\textwidth]{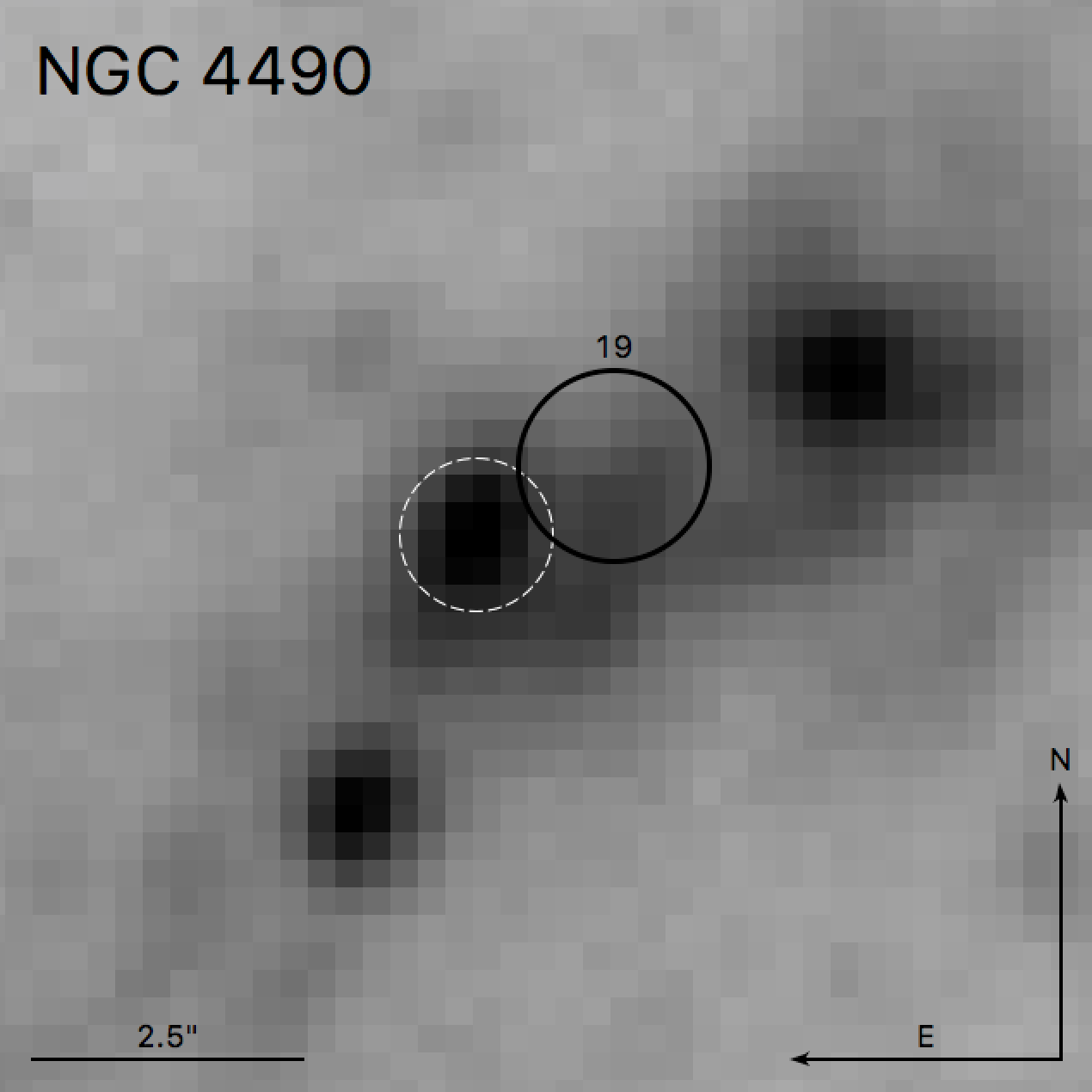}%
        \subcaption{NGC 4490, ULX ID 19, 3-$\sigma$ = 0.9"}
        \label{fig:4490ulx19}
         \vspace{0.5cm}                
    \end{minipage}
    \begin{minipage}{0.333\textwidth}
        \includegraphics[width=\textwidth]{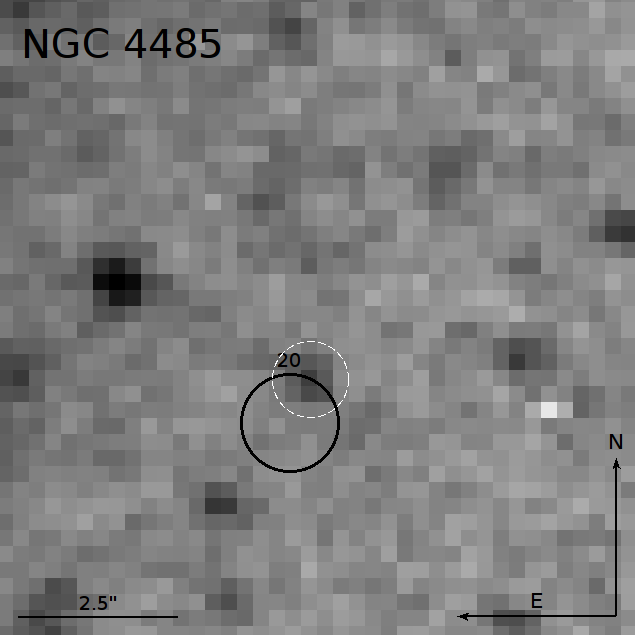}%
        \subcaption{NGC 4485, ULX ID 20, 3-$\sigma$ = 0.8"}
        \label{fig:4485ulx20}
         \vspace{0.5cm}                
    \end{minipage}%
    \begin{minipage}{0.333\textwidth}
        \includegraphics[width=\textwidth]{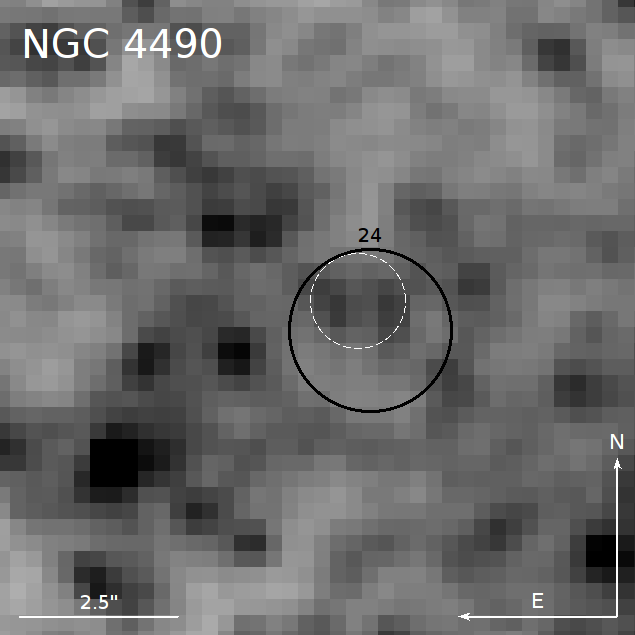}%
         \subcaption{NGC 4490, ULX ID 24, 3-$\sigma$ = 1.3"}
         \label{fig:4490ulx24}
         \vspace{0.5cm}                
    \end{minipage}%
    \begin{minipage}{0.333\textwidth}
        \includegraphics[width=\textwidth]{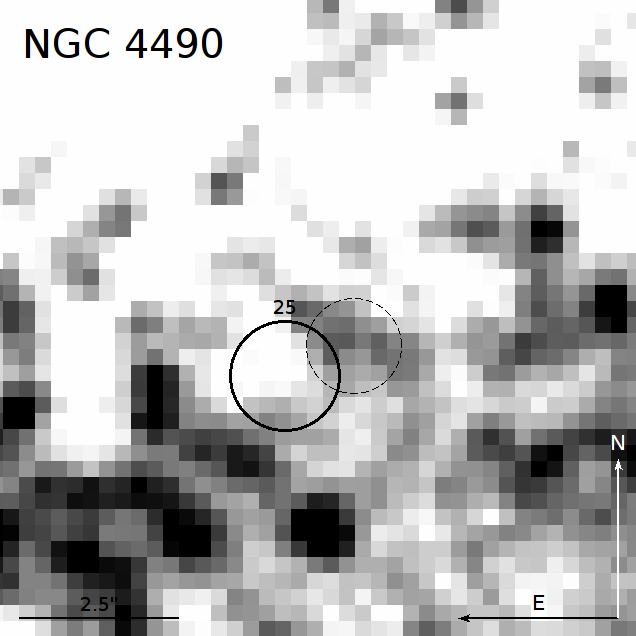}%
         \subcaption{NGC 4490, ULX ID 25, 3-$\sigma$ = 0.9"}
         \label{fig:4490ulx25}
          \vspace{0.5cm}              
    \end{minipage} 
	\caption{Finder charts of the ULXs with a NIR candidate counterpart. The black/white solid circles correspond to the 99.7$\%$ confidence radius around the position of the ULXs, whereas the black/white dashed circles mark the candidate counterpart as detected with {\scshape SExtractor}. Each image provides the value for the 99.7$\%$ (3-$\sigma$) uncertainty radius. It is important to note that some faint sources seen by eye are not significantly detected by {\scshape SExtractor} (e.g. Figures~\ref{fig:manycandidates},~\ref{fig:2403ulx9},~\ref{fig:re4258},~\ref{fig:4490ulx24}).}
    \label{fig:images}  
\end{figure*} 
  
\begin{figure*}
	\ContinuedFloat
	    \begin{minipage}{0.333\textwidth}
        \includegraphics[width=\textwidth]{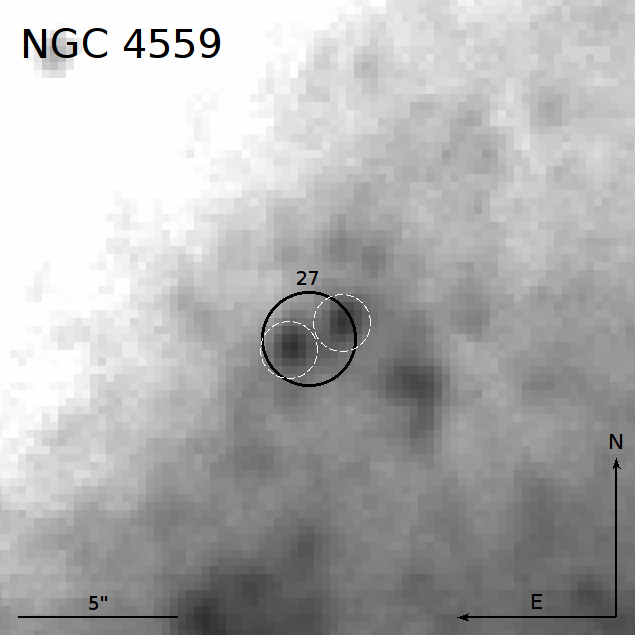}%
        \subcaption{NGC 4559, ULX ID 27, 3-$\sigma$ = 1.5"}  
        \label{fig:4559ulx27}
         \vspace{0.5cm}              
    \end{minipage}%
    \begin{minipage}{0.333\textwidth}
        \includegraphics[width=\textwidth]{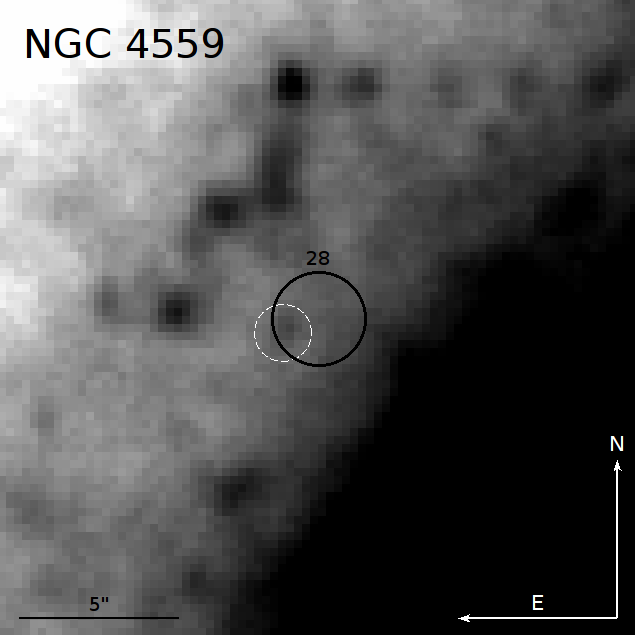}%
         \subcaption{NGC 4559, ULX ID 28, 3-$\sigma$ = 1.5"}
         \label{fig:4559ulx28}  
         \vspace{0.5cm}              
    \end{minipage}%
    \begin{minipage}{0.333\textwidth}
        \includegraphics[width=\textwidth]{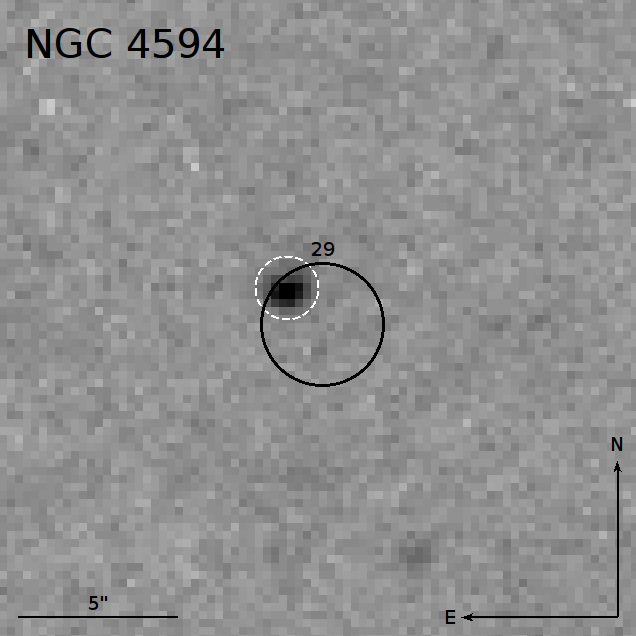}%
        \subcaption{NGC 4594, ULX ID 29, 3-$\sigma$ = 1.9"}
        \label{fig:4594ulx29} 
         \vspace{0.5cm}               
    \end{minipage}
    \begin{minipage}{0.333\textwidth}
        \includegraphics[width=\textwidth]{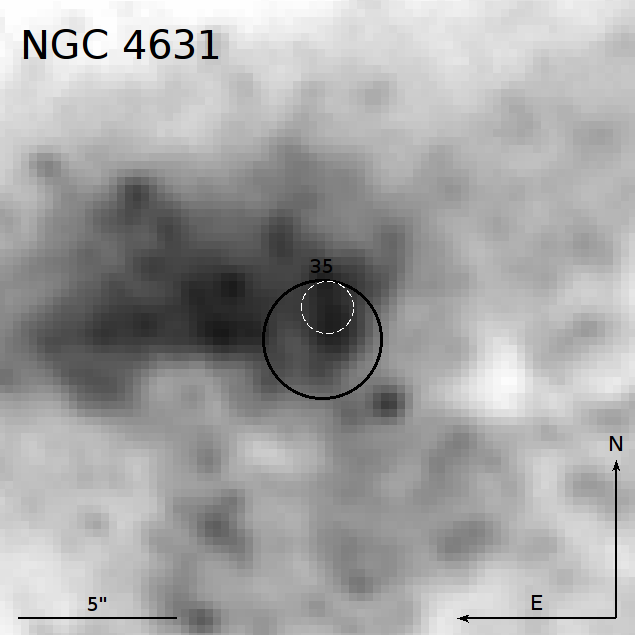}%
         \subcaption{NGC 4631, ULX ID 35, 3-$\sigma$ = 1.8"} 
         \label{fig:4631ulx35}
         \vspace{0.5cm}               
    \end{minipage}%
    \begin{minipage}{0.333\textwidth}
        \includegraphics[width=\textwidth]{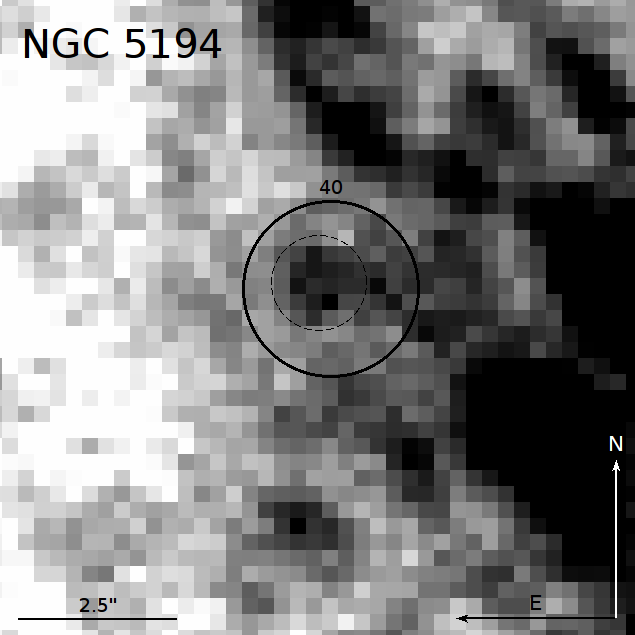}%
        \subcaption{NGC 5194, ULX ID 40, 3-$\sigma$ = 1.4"}
        \label{fig:5194ulx40}
         \vspace{0.5cm}                
    \end{minipage}%
    \begin{minipage}{0.333\textwidth}
        \includegraphics[width=\textwidth]{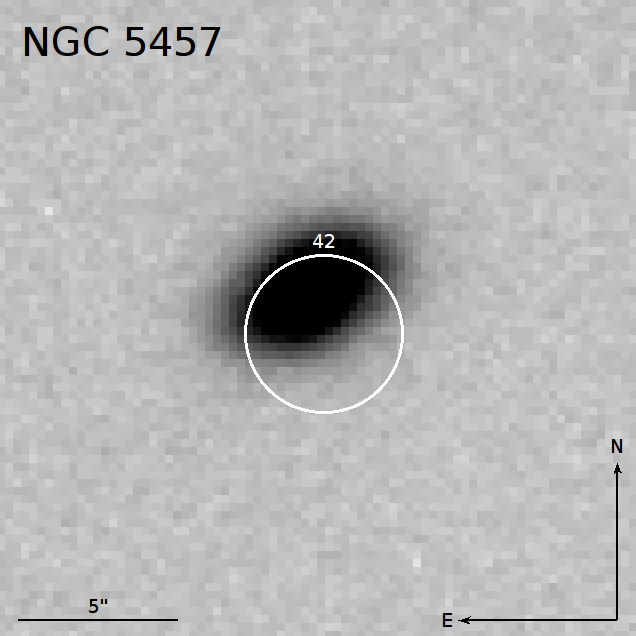}%
        \subcaption{NGC 5457, ULX ID 42, 3-$\sigma$ = 2.5"}
        \label{fig:5457ulx42}
         \vspace{0.5cm}                     
    \end{minipage} 
     \begin{minipage}{0.333\textwidth}
        \includegraphics[width=\textwidth]{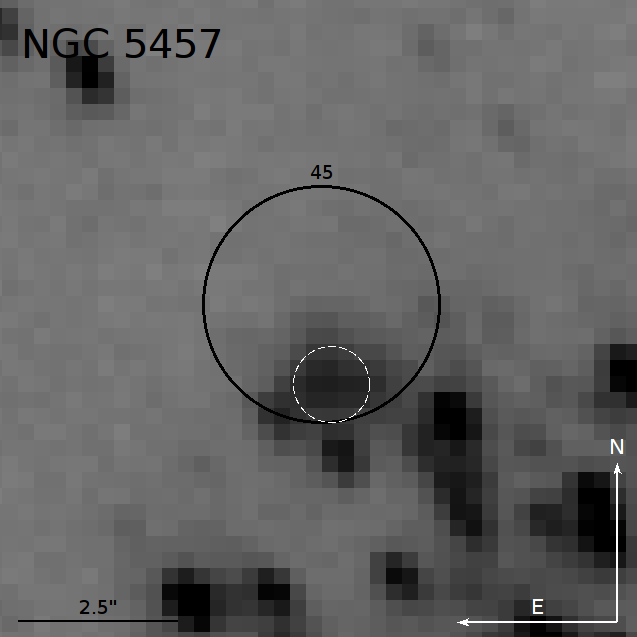}%
        \subcaption{NGC 5457, ULX ID 45, 3-$\sigma$ = 1.9"}
        \label{fig:5457ulx45}
          \vspace{0.5cm}               
    \end{minipage}%
\caption{--- {\it continued}.}
\end{figure*}




\bibliographystyle{mnras}
\bibliography{library}

\begin{thebibliography}{}
\makeatletter
\relax
\def\mn@urlcharsother{\let\do\@makeother \do\$\do\&\do\#\do\^\do\_\do\%\do\~}
\def\mn@doi{\begingroup\mn@urlcharsother \@ifnextchar [ {\mn@doi@}
  {\mn@doi@[]}}
\def\mn@doi@[#1]#2{\def\@tempa{#1}\ifx\@tempa\@empty \href
  {http://dx.doi.org/#2} {doi:#2}\else \href {http://dx.doi.org/#2} {#1}\fi
  \endgroup}
\def\mn@eprint#1#2{\mn@eprint@#1:#2::\@nil}
\def\mn@eprint@arXiv#1{\href {http://arxiv.org/abs/#1} {{\tt arXiv:#1}}}
\def\mn@eprint@dblp#1{\href {http://dblp.uni-trier.de/rec/bibtex/#1.xml}
  {dblp:#1}}
\def\mn@eprint@#1:#2:#3:#4\@nil{\def\@tempa {#1}\def\@tempb {#2}\def\@tempc
  {#3}\ifx \@tempc \@empty \let \@tempc \@tempb \let \@tempb \@tempa \fi \ifx
  \@tempb \@empty \def\@tempb {arXiv}\fi \@ifundefined
  {mn@eprint@\@tempb}{\@tempb:\@tempc}{\expandafter \expandafter \csname
  mn@eprint@\@tempb\endcsname \expandafter{\@tempc}}}

\bibitem[\protect\citeauthoryear{{Abbott} et~al.,}{{Abbott}
  et~al.}{2016}]{2016PhRvL.116f1102A}
{Abbott} B.~P.,  et~al., 2016, \mn@doi [Physical Review Letters]
  {10.1103/PhysRevLett.116.061102}, \href
  {http://adsabs.harvard.edu/abs/2016PhRvL.116f1102A} {116, 061102}

\bibitem[\protect\citeauthoryear{{Ba{\~n}ados} et~al.,}{{Ba{\~n}ados}
  et~al.}{2016}]{2016arXiv160803279B}
{Ba{\~n}ados} E.,  et~al., 2016, preprint, \href
  {http://adsabs.harvard.edu/abs/2016arXiv160803279B} {} (\mn@eprint {arXiv}
  {1608.03279})

\bibitem[\protect\citeauthoryear{{Bachetti} et~al.,}{{Bachetti}
  et~al.}{2014}]{2014Natur.514..202B}
{Bachetti} M.,  et~al., 2014, \mn@doi [\nat] {10.1038/nature13791}, \href
  {http://adsabs.harvard.edu/abs/2014Natur.514..202B} {514, 202}

\bibitem[\protect\citeauthoryear{{Bauer}, {Brandt}  \& {Lehmer}}{{Bauer}
  et~al.}{2003}]{2003AJ....126.2797B}
{Bauer} F.~E.,  {Brandt} W.~N.,   {Lehmer} B.,  2003, \mn@doi [\aj]
  {10.1086/379139}, \href {http://adsabs.harvard.edu/abs/2003AJ....126.2797B}
  {126, 2797}

\bibitem[\protect\citeauthoryear{{Begelman}}{{Begelman}}{2002}]{2002ApJ...568L..97B}
{Begelman} M.~C.,  2002, \mn@doi [\apjl] {10.1086/340457}, \href
  {http://adsabs.harvard.edu/abs/2002ApJ...568L..97B} {568, L97}

\bibitem[\protect\citeauthoryear{{Bertin}}{{Bertin}}{2006}]{2006ASPC..351..112B}
{Bertin} E.,  2006, in {Gabriel} C.,  {Arviset} C.,  {Ponz} D.,   {Enrique} S.,
   eds,  Astronomical Society of the Pacific Conference Series Vol. 351,
  Astronomical Data Analysis Software and Systems XV. p.~112

\bibitem[\protect\citeauthoryear{{Bertin} \& {Arnouts}}{{Bertin} \&
  {Arnouts}}{1996}]{1996A&AS..117..393B}
{Bertin} E.,  {Arnouts} S.,  1996, \mn@doi [\aaps] {10.1051/aas:1996164}, \href
  {http://adsabs.harvard.edu/abs/1996A%26AS..117..393B} {117, 393}

\bibitem[\protect\citeauthoryear{{Bertin}, {Mellier}, {Radovich}, {Missonnier},
  {Didelon}  \& {Morin}}{{Bertin} et~al.}{2002}]{2002ASPC..281..228B}
{Bertin} E.,  {Mellier} Y.,  {Radovich} M.,  {Missonnier} G.,  {Didelon} P.,
  {Morin} B.,  2002, in {Bohlender} D.~A.,  {Durand} D.,   {Handley} T.~H.,
  eds,  Astronomical Society of the Pacific Conference Series Vol. 281,
  Astronomical Data Analysis Software and Systems XI. p.~228

\bibitem[\protect\citeauthoryear{{Casares} \& {Jonker}}{{Casares} \&
  {Jonker}}{2014}]{2014SSRv..183..223C}
{Casares} J.,  {Jonker} P.~G.,  2014, \mn@doi [\ssr]
  {10.1007/s11214-013-0030-6}, \href
  {http://adsabs.harvard.edu/abs/2014SSRv..183..223C} {183, 223}

\bibitem[\protect\citeauthoryear{{Colbert} \& {Miller}}{{Colbert} \&
  {Miller}}{2006}]{2006tmgm.meet..530C}
{Colbert} E.~J.~M.,  {Miller} M.~C.,  2006, in {Novello} M.,  {Perez
  Bergliaffa} S.,   {Ruffini} R.,  eds, The Tenth Marcel Grossmann Meeting.
  Proceedings of the MG10 Meeting held at Brazilian Center for Research in
  Physics (CBPF), Rio de Janeiro, Brazil, 20-26 July 2003, Eds.: M{\'a}rio
  Novello; Santiago Perez Bergliaffa; Remo Ruffini. Singapore: World Scientific
  Publishing, in 3 volumes, ISBN 981-256-667-8 (set), ISBN 981-256-980-4 (Part
  A), ISBN 981-256-979-0 (Part B), ISBN 981-256-978-2 (Part C), 2006, XLVIII +
  2492 pp.: 2006, p.530. p.~530 (\mn@eprint {} {astro-ph/0402677}),
  \mn@doi{10.1142/9789812704030_0032}

\bibitem[\protect\citeauthoryear{{Copperwheat}, {Cropper}, {Soria}  \&
  {Wu}}{{Copperwheat} et~al.}{2005}]{2005MNRAS.362...79C}
{Copperwheat} C.,  {Cropper} M.,  {Soria} R.,   {Wu} K.,  2005, \mn@doi
  [\mnras] {10.1111/j.1365-2966.2005.09223.x}, \href
  {http://adsabs.harvard.edu/abs/2005MNRAS.362...79C} {362, 79}

\bibitem[\protect\citeauthoryear{{Copperwheat}, {Cropper}, {Soria}  \&
  {Wu}}{{Copperwheat} et~al.}{2007}]{2007MNRAS.376.1407C}
{Copperwheat} C.,  {Cropper} M.,  {Soria} R.,   {Wu} K.,  2007, \mn@doi
  [\mnras] {10.1111/j.1365-2966.2007.11551.x}, \href
  {http://adsabs.harvard.edu/abs/2007MNRAS.376.1407C} {376, 1407}

\bibitem[\protect\citeauthoryear{{Cropper}, {Soria}, {Mushotzky}, {Wu},
  {Markwardt}  \& {Pakull}}{{Cropper} et~al.}{2004}]{2004MNRAS.349...39C}
{Cropper} M.,  {Soria} R.,  {Mushotzky} R.~F.,  {Wu} K.,  {Markwardt} C.~B.,
  {Pakull} M.,  2004, \mn@doi [\mnras] {10.1111/j.1365-2966.2004.07480.x},
  \href {http://adsabs.harvard.edu/abs/2004MNRAS.349...39C} {349, 39}

\bibitem[\protect\citeauthoryear{{D'Abrusco}, {Massaro}, {Ajello}, {Grindlay},
  {Smith}  \& {Tosti}}{{D'Abrusco} et~al.}{2012}]{2012ApJ...748...68D}
{D'Abrusco} R.,  {Massaro} F.,  {Ajello} M.,  {Grindlay} J.~E.,  {Smith} H.~A.,
    {Tosti} G.,  2012, \mn@doi [\apj] {10.1088/0004-637X/748/1/68}, \href
  {http://adsabs.harvard.edu/abs/2012ApJ...748...68D} {748, 68}

\bibitem[\protect\citeauthoryear{{Dobrzycki}, {Ebeling}, {Glotfelty},
  {Freeman}, {Damiani}, {Elvis}  \& {Calderwood}}{{Dobrzycki}
  et~al.}{1999}]{ciao}
{Dobrzycki} A.,  {Ebeling} H.,  {Glotfelty} K.,  {Freeman} P.,  {Damiani} F.,
  {Elvis} M.,   {Calderwood} T.,  1999. p.~142

\bibitem[\protect\citeauthoryear{{Drilling} \& {Landolt}}{{Drilling} \&
  {Landolt}}{2000}]{2000asqu.book..381D}
{Drilling} J.~S.,  {Landolt} A.~U.,  2000, {Normal Stars}.
p.~381

\bibitem[\protect\citeauthoryear{{Ebisuzaki} et~al.,}{{Ebisuzaki}
  et~al.}{2001}]{2001ApJ...562L..19E}
{Ebisuzaki} T.,  et~al., 2001, \mn@doi [\apjl] {10.1086/338118}, \href
  {http://adsabs.harvard.edu/abs/2001ApJ...562L..19E} {562, L19}

\bibitem[\protect\citeauthoryear{{Elias}, {Frogel}  \& {Humphreys}}{{Elias}
  et~al.}{1985}]{1985ApJS...57...91E}
{Elias} J.~H.,  {Frogel} J.~A.,   {Humphreys} R.~M.,  1985, \mn@doi [\apjs]
  {10.1086/190997}, \href {http://adsabs.harvard.edu/abs/1985ApJS...57...91E}
  {57, 91}

\bibitem[\protect\citeauthoryear{{Evans} et~al.,}{{Evans}
  et~al.}{2010}]{2010ApJS..189...37E}
{Evans} I.~N.,  et~al., 2010, \mn@doi [\apjs] {10.1088/0067-0049/189/1/37},
  \href {http://adsabs.harvard.edu/abs/2010ApJS..189...37E} {189, 37}

\bibitem[\protect\citeauthoryear{{Fabbiano}, {Zezas}  \& {Murray}}{{Fabbiano}
  et~al.}{2001}]{2001ApJ...554.1035F}
{Fabbiano} G.,  {Zezas} A.,   {Murray} S.~S.,  2001, \mn@doi [\apj]
  {10.1086/321397}, \href {http://adsabs.harvard.edu/abs/2001ApJ...554.1035F}
  {554, 1035}

\bibitem[\protect\citeauthoryear{{Fabrika}, {Ueda}, {Vinokurov}, {Sholukhova}
  \& {Shidatsu}}{{Fabrika} et~al.}{2015}]{2015NatPh..11..551F}
{Fabrika} S.,  {Ueda} Y.,  {Vinokurov} A.,  {Sholukhova} O.,   {Shidatsu} M.,
  2015, \mn@doi [Nature Physics] {10.1038/nphys3348}, \href
  {http://cdsads.u-strasbg.fr/abs/2015NatPh..11..551F} {11, 551}

\bibitem[\protect\citeauthoryear{{Fan} et~al.,}{{Fan}
  et~al.}{2000}]{2000AJ....120.1167F}
{Fan} X.,  et~al., 2000, \mn@doi [\aj] {10.1086/301534}, \href
  {http://adsabs.harvard.edu/abs/2000AJ....120.1167F} {120, 1167}

\bibitem[\protect\citeauthoryear{{Feng} \& {Soria}}{{Feng} \&
  {Soria}}{2011}]{2011NewAR..55..166F}
{Feng} H.,  {Soria} R.,  2011, \mn@doi [\nar] {10.1016/j.newar.2011.08.002},
  \href {http://adsabs.harvard.edu/abs/2011NewAR..55..166F} {55, 166}

\bibitem[\protect\citeauthoryear{{Fruscione} et~al.,}{{Fruscione}
  et~al.}{2006}]{2006SPIE.6270E..1VF}
{Fruscione} A.,  et~al., 2006, in Society of Photo-Optical Instrumentation
  Engineers (SPIE) Conference Series. p. 62701V, \mn@doi{10.1117/12.671760}

\bibitem[\protect\citeauthoryear{{Gao}, {Wang}, {Appleton}  \& {Lucas}}{{Gao}
  et~al.}{2003}]{2003ApJ...596L.171G}
{Gao} Y.,  {Wang} Q.~D.,  {Appleton} P.~N.,   {Lucas} R.~A.,  2003, \mn@doi
  [\apjl] {10.1086/379598}, \href
  {http://adsabs.harvard.edu/abs/2003ApJ...596L.171G} {596, L171}

\bibitem[\protect\citeauthoryear{{Gladstone}, {Roberts}  \& {Done}}{{Gladstone}
  et~al.}{2009}]{2009MNRAS.397.1836G}
{Gladstone} J.~C.,  {Roberts} T.~P.,   {Done} C.,  2009, \mn@doi [\mnras]
  {10.1111/j.1365-2966.2009.15123.x}, \href
  {http://adsabs.harvard.edu/abs/2009MNRAS.397.1836G} {397, 1836}

\bibitem[\protect\citeauthoryear{{Gladstone}, {Copperwheat}, {Heinke},
  {Roberts}, {Cartwright}, {Levan}  \& {Goad}}{{Gladstone}
  et~al.}{2013}]{2013ApJS..206...14G}
{Gladstone} J.~C.,  {Copperwheat} C.,  {Heinke} C.~O.,  {Roberts} T.~P.,
  {Cartwright} T.~F.,  {Levan} A.~J.,   {Goad} M.~R.,  2013, \mn@doi [\apjs]
  {10.1088/0067-0049/206/2/14}, \href
  {http://adsabs.harvard.edu/abs/2013ApJS..206...14G} {206, 14}

\bibitem[\protect\citeauthoryear{{Guti{\'e}rrez}}{{Guti{\'e}rrez}}{2013}]{2013A&A...549A..81G}
{Guti{\'e}rrez} C.~M.,  2013, \mn@doi [\aap] {10.1051/0004-6361/201220433},
  \href {http://adsabs.harvard.edu/abs/2013A%26A...549A..81G} {549, A81}

\bibitem[\protect\citeauthoryear{{Guti{\'e}rrez} \&
  {L{\'o}pez-Corredoira}}{{Guti{\'e}rrez} \&
  {L{\'o}pez-Corredoira}}{2006}]{2006IAUS..230..310G}
{Guti{\'e}rrez} C.~M.,  {L{\'o}pez-Corredoira} M.,  2006, in {Meurs} E.~J.~A.,
  {Fabbiano} G.,  eds,  IAU Symposium Vol. 230, Populations of High Energy
  Sources in Galaxies. pp 310--311, \mn@doi{10.1017/S1743921306008556}

\bibitem[\protect\citeauthoryear{{Heida} et~al.,}{{Heida}
  et~al.}{2014}]{2014MNRAS.442.1054H}
{Heida} M.,  et~al., 2014, \mn@doi [\mnras] {10.1093/mnras/stu928}, \href
  {http://adsabs.harvard.edu/abs/2014MNRAS.442.1054H} {442, 1054}

\bibitem[\protect\citeauthoryear{{Heida} et~al.,}{{Heida}
  et~al.}{2015}]{2015MNRAS.453.3510H}
{Heida} M.,  et~al., 2015, \mn@doi [\mnras] {10.1093/mnras/stv1853}, \href
  {http://adsabs.harvard.edu/abs/2015MNRAS.453.3510H} {453, 3510}

\bibitem[\protect\citeauthoryear{{Heida}, {Jonker}, {Torres}, {Roberts},
  {Walton}, {Moon}, {Stern}  \& {Harrison}}{{Heida}
  et~al.}{2016}]{2016MNRAS.459..771H}
{Heida} M.,  {Jonker} P.~G.,  {Torres} M.~A.~P.,  {Roberts} T.~P.,  {Walton}
  D.~J.,  {Moon} D.-S.,  {Stern} D.,   {Harrison} F.~A.,  2016, \mn@doi
  [\mnras] {10.1093/mnras/stw695}, \href
  {http://adsabs.harvard.edu/abs/2016MNRAS.459..771H} {459, 771}

\bibitem[\protect\citeauthoryear{{Hwang} \& {Lee}}{{Hwang} \&
  {Lee}}{2008}]{2008AJ....135.1567H}
{Hwang} N.,  {Lee} M.~G.,  2008, \mn@doi [\aj] {10.1088/0004-6256/135/4/1567},
  \href {http://cdsads.u-strasbg.fr/abs/2008AJ....135.1567H} {135, 1567}

\bibitem[\protect\citeauthoryear{{Israel} et~al.,}{{Israel}
  et~al.}{2016}]{2016arXiv160907375I}
{Israel} G.~L.,  et~al., 2016, preprint, \href
  {http://adsabs.harvard.edu/abs/2016arXiv160907375I} {} (\mn@eprint {arXiv}
  {1609.07375})

\bibitem[\protect\citeauthoryear{{Israel} et~al.,}{{Israel}
  et~al.}{2017}]{2017MNRAS.466L..48I}
{Israel} G.~L.,  et~al., 2017, \mn@doi [\mnras] {10.1093/mnrasl/slw218}, \href
  {http://adsabs.harvard.edu/abs/2017MNRAS.466L..48I} {466, L48}

\bibitem[\protect\citeauthoryear{{Jonker}, {Torres}, {Fabian}, {Heida},
  {Miniutti}  \& {Pooley}}{{Jonker} et~al.}{2010}]{2010MNRAS.407..645J}
{Jonker} P.~G.,  {Torres} M.~A.~P.,  {Fabian} A.~C.,  {Heida} M.,  {Miniutti}
  G.,   {Pooley} D.,  2010, \mn@doi [\mnras]
  {10.1111/j.1365-2966.2010.16943.x}, \href
  {http://adsabs.harvard.edu/abs/2010MNRAS.407..645J} {407, 645}

\bibitem[\protect\citeauthoryear{{Jonker} et~al.,}{{Jonker}
  et~al.}{2012}]{2012ApJ...758...28J}
{Jonker} P.~G.,  et~al., 2012, \mn@doi [\apj] {10.1088/0004-637X/758/1/28},
  \href {http://adsabs.harvard.edu/abs/2012ApJ...758...28J} {758, 28}

\bibitem[\protect\citeauthoryear{{Kalogera} \& {Baym}}{{Kalogera} \&
  {Baym}}{1996}]{1996ApJ...470L..61K}
{Kalogera} V.,  {Baym} G.,  1996, \mn@doi [\apjl] {10.1086/310296}, \href
  {http://adsabs.harvard.edu/abs/1996ApJ...470L..61K} {470, L61}

\bibitem[\protect\citeauthoryear{{King}, {Davies}, {Ward}, {Fabbiano}  \&
  {Elvis}}{{King} et~al.}{2001}]{2001ApJ...552L.109K}
{King} A.~R.,  {Davies} M.~B.,  {Ward} M.~J.,  {Fabbiano} G.,   {Elvis} M.,
  2001, \mn@doi [\apjl] {10.1086/320343}, \href
  {http://adsabs.harvard.edu/abs/2001ApJ...552L.109K} {552, L109}

\bibitem[\protect\citeauthoryear{{Kong}}{{Kong}}{2003}]{2003MNRAS.346..265K}
{Kong} A.~K.~H.,  2003, \mn@doi [\mnras] {10.1046/j.1365-2966.2003.07086.x},
  \href {http://adsabs.harvard.edu/abs/2003MNRAS.346..265K} {346, 265}

\bibitem[\protect\citeauthoryear{{Lin}, {Webb}  \& {Barret}}{{Lin}
  et~al.}{2012}]{2012ApJ...756...27L}
{Lin} D.,  {Webb} N.~A.,   {Barret} D.,  2012, \mn@doi [\apj]
  {10.1088/0004-637X/756/1/27}, \href
  {http://adsabs.harvard.edu/abs/2012ApJ...756...27L} {756, 27}

\bibitem[\protect\citeauthoryear{{Liu}}{{Liu}}{2011}]{2011ApJS..192...10L}
{Liu} J.,  2011, \mn@doi [\apjs] {10.1088/0067-0049/192/1/10}, \href
  {http://adsabs.harvard.edu/abs/2011ApJS..192...10L} {192, 10}

\bibitem[\protect\citeauthoryear{{Liu} \& {Bregman}}{{Liu} \&
  {Bregman}}{2005}]{2005ApJS..157...59L}
{Liu} J.-F.,  {Bregman} J.~N.,  2005, \mn@doi [\apjs] {10.1086/427170}, \href
  {http://adsabs.harvard.edu/abs/2005ApJS..157...59L} {157, 59}

\bibitem[\protect\citeauthoryear{{Liu} \& {Mirabel}}{{Liu} \&
  {Mirabel}}{2005}]{2005A&A...429.1125L}
{Liu} Q.~Z.,  {Mirabel} I.~F.,  2005, \mn@doi [\aap]
  {10.1051/0004-6361:20041878}, \href
  {http://cdsads.u-strasbg.fr/abs/2005A%26A...429.1125L} {429, 1125}

\bibitem[\protect\citeauthoryear{{Liu}, {Bregman}  \& {Seitzer}}{{Liu}
  et~al.}{2002}]{2002ApJ...580L..31L}
{Liu} J.-F.,  {Bregman} J.~N.,   {Seitzer} P.,  2002, \mn@doi [\apjl]
  {10.1086/345420}, \href {http://adsabs.harvard.edu/abs/2002ApJ...580L..31L}
  {580, L31}

\bibitem[\protect\citeauthoryear{{Monet} et~al.,}{{Monet}
  et~al.}{2003}]{2003AJ....125..984M}
{Monet} D.~G.,  et~al., 2003, \mn@doi [\aj] {10.1086/345888}, \href
  {http://adsabs.harvard.edu/abs/2003AJ....125..984M} {125, 984}

\bibitem[\protect\citeauthoryear{{Moon}, {Eikenberry}  \& {Wasserman}}{{Moon}
  et~al.}{2003}]{2003ApJ...586.1280M}
{Moon} D.-S.,  {Eikenberry} S.~S.,   {Wasserman} I.~M.,  2003, \mn@doi [\apj]
  {10.1086/367826}, \href {http://adsabs.harvard.edu/abs/2003ApJ...586.1280M}
  {586, 1280}

\bibitem[\protect\citeauthoryear{{Motch}, {Pakull}, {Soria}, {Gris{\'e}}  \&
  {Pietrzy{\'n}ski}}{{Motch} et~al.}{2014}]{2014Natur.514..198M}
{Motch} C.,  {Pakull} M.~W.,  {Soria} R.,  {Gris{\'e}} F.,   {Pietrzy{\'n}ski}
  G.,  2014, \mn@doi [\nat] {10.1038/nature13730}, \href
  {http://adsabs.harvard.edu/abs/2014Natur.514..198M} {514, 198}

\bibitem[\protect\citeauthoryear{{Orosz} et~al.,}{{Orosz}
  et~al.}{2007}]{2007Natur.449..872O}
{Orosz} J.~A.,  et~al., 2007, \mn@doi [\nat] {10.1038/nature06218}, \href
  {http://cdsads.u-strasbg.fr/abs/2007Natur.449..872O} {449, 872}

\bibitem[\protect\citeauthoryear{{Patruno} \& {Zampieri}}{{Patruno} \&
  {Zampieri}}{2008}]{2008MNRAS.386..543P}
{Patruno} A.,  {Zampieri} L.,  2008, \mn@doi [\mnras]
  {10.1111/j.1365-2966.2008.13063.x}, \href
  {http://adsabs.harvard.edu/abs/2008MNRAS.386..543P} {386, 543}

\bibitem[\protect\citeauthoryear{{P{\'e}rez-Ram{\'{\i}}rez}, {Mezcua}, {Leon}
  \& {Caballero-Garc{\'{\i}}a}}{{P{\'e}rez-Ram{\'{\i}}rez}
  et~al.}{2011}]{2011AN....332..384P}
{P{\'e}rez-Ram{\'{\i}}rez} D.,  {Mezcua} M.,  {Leon} S.,
  {Caballero-Garc{\'{\i}}a} M.~D.,  2011, \mn@doi [Astronomische Nachrichten]
  {10.1002/asna.201011505}, \href
  {http://adsabs.harvard.edu/abs/2011AN....332..384P} {332, 384}

\bibitem[\protect\citeauthoryear{{Pier}, {Munn}, {Hindsley}, {Hennessy},
  {Kent}, {Lupton}  \& {Ivezi{\'c}}}{{Pier} et~al.}{2003}]{2003AJ....125.1559P}
{Pier} J.~R.,  {Munn} J.~A.,  {Hindsley} R.~B.,  {Hennessy} G.~S.,  {Kent}
  S.~M.,  {Lupton} R.~H.,   {Ivezi{\'c}} {\v Z}.,  2003, \mn@doi [\aj]
  {10.1086/346138}, \href {http://adsabs.harvard.edu/abs/2003AJ....125.1559P}
  {125, 1559}

\bibitem[\protect\citeauthoryear{{Poutanen}, {Fabrika}, {Valeev}, {Sholukhova}
  \& {Greiner}}{{Poutanen} et~al.}{2013}]{2013MNRAS.432..506P}
{Poutanen} J.,  {Fabrika} S.,  {Valeev} A.~F.,  {Sholukhova} O.,   {Greiner}
  J.,  2013, \mn@doi [\mnras] {10.1093/mnras/stt487}, \href
  {http://adsabs.harvard.edu/abs/2013MNRAS.432..506P} {432, 506}

\bibitem[\protect\citeauthoryear{{Ptak}, {Colbert}, {van der Marel}, {Roye},
  {Heckman}  \& {Towne}}{{Ptak} et~al.}{2006}]{2006ApJS..166..154P}
{Ptak} A.,  {Colbert} E.,  {van der Marel} R.~P.,  {Roye} E.,  {Heckman} T.,
  {Towne} B.,  2006, \mn@doi [\apjs] {10.1086/505218}, \href
  {http://adsabs.harvard.edu/abs/2006ApJS..166..154P} {166, 154}

\bibitem[\protect\citeauthoryear{{Roberts}, {Warwick}, {Ward}  \&
  {Murray}}{{Roberts} et~al.}{2002}]{2002MNRAS.337..677R}
{Roberts} T.~P.,  {Warwick} R.~S.,  {Ward} M.~J.,   {Murray} S.~S.,  2002,
  \mn@doi [\mnras] {10.1046/j.1365-8711.2002.05950.x}, \href
  {http://cdsads.u-strasbg.fr/abs/2002MNRAS.337..677R} {337, 677}

\bibitem[\protect\citeauthoryear{{Roberts}, {Gladstone}, {Goulding},
  {Swinbank}, {Ward}, {Goad}  \& {Levan}}{{Roberts}
  et~al.}{2011}]{2011AN....332..398R}
{Roberts} T.~P.,  {Gladstone} J.~C.,  {Goulding} A.~D.,  {Swinbank} A.~M.,
  {Ward} M.~J.,  {Goad} M.~R.,   {Levan} A.~J.,  2011, \mn@doi [Astronomische
  Nachrichten] {10.1002/asna.201011508}, \href
  {http://cdsads.u-strasbg.fr/abs/2011AN....332..398R} {332, 398}

\bibitem[\protect\citeauthoryear{{Roeser}, {Demleitner}  \&
  {Schilbach}}{{Roeser} et~al.}{2010}]{2010AJ....139.2440R}
{Roeser} S.,  {Demleitner} M.,   {Schilbach} E.,  2010, \mn@doi [\aj]
  {10.1088/0004-6256/139/6/2440}, \href
  {http://ads.ari.uni-heidelberg.de/abs/2010AJ....139.2440R} {139, 2440}

\bibitem[\protect\citeauthoryear{{Schirmer}}{{Schirmer}}{2013}]{2013ApJS..209...21S}
{Schirmer} M.,  2013, \mn@doi [\apjs] {10.1088/0067-0049/209/2/21}, \href
  {http://adsabs.harvard.edu/abs/2013ApJS..209...21S} {209, 21}

\bibitem[\protect\citeauthoryear{{Schlegel} \& {Pannuti}}{{Schlegel} \&
  {Pannuti}}{2003}]{2003AJ....125.3025S}
{Schlegel} E.~M.,  {Pannuti} T.~G.,  2003, \mn@doi [\aj] {10.1086/374990},
  \href {http://adsabs.harvard.edu/abs/2003AJ....125.3025S} {125, 3025}

\bibitem[\protect\citeauthoryear{{Secrest} et~al.,}{{Secrest}
  et~al.}{2015}]{2015ApJ...798...38S}
{Secrest} N.~J.,  et~al., 2015, \mn@doi [\apj] {10.1088/0004-637X/798/1/38},
  \href {http://adsabs.harvard.edu/abs/2015ApJ...798...38S} {798, 38}

\bibitem[\protect\citeauthoryear{{Skrutskie} et~al.,}{{Skrutskie}
  et~al.}{2006}]{2006AJ....131.1163S}
{Skrutskie} M.~F.,  et~al., 2006, \mn@doi [\aj] {10.1086/498708}, \href
  {http://adsabs.harvard.edu/abs/2006AJ....131.1163S} {131, 1163}

\bibitem[\protect\citeauthoryear{{Sutton}, {Roberts}, {Walton}, {Gladstone}  \&
  {Scott}}{{Sutton} et~al.}{2012}]{2012MNRAS.423.1154S}
{Sutton} A.~D.,  {Roberts} T.~P.,  {Walton} D.~J.,  {Gladstone} J.~C.,
  {Scott} A.~E.,  2012, \mn@doi [\mnras] {10.1111/j.1365-2966.2012.20944.x},
  \href {http://adsabs.harvard.edu/abs/2012MNRAS.423.1154S} {423, 1154}

\bibitem[\protect\citeauthoryear{{Swartz}, {Ghosh}, {Tennant}  \&
  {Wu}}{{Swartz} et~al.}{2004}]{2004ApJS..154..519S}
{Swartz} D.~A.,  {Ghosh} K.~K.,  {Tennant} A.~F.,   {Wu} K.,  2004, \mn@doi
  [\apjs] {10.1086/422842}, \href
  {http://cdsads.u-strasbg.fr/abs/2004ApJS..154..519S} {154, 519}

\bibitem[\protect\citeauthoryear{{Swartz}, {Soria}, {Tennant}  \&
  {Yukita}}{{Swartz} et~al.}{2011}]{2011ApJ...741...49S}
{Swartz} D.~A.,  {Soria} R.,  {Tennant} A.~F.,   {Yukita} M.,  2011, \mn@doi
  [\apj] {10.1088/0004-637X/741/1/49}, \href
  {http://adsabs.harvard.edu/abs/2011ApJ...741...49S} {741, 49}

\bibitem[\protect\citeauthoryear{{Tikhonov}, {Lebedev}  \&
  {Galazutdinova}}{{Tikhonov} et~al.}{2015}]{2015AstL...41..239T}
{Tikhonov} N.~A.,  {Lebedev} V.~S.,   {Galazutdinova} O.~A.,  2015, \mn@doi
  [Astronomy Letters] {10.1134/S1063773715060080}, \href
  {http://adsabs.harvard.edu/abs/2015AstL...41..239T} {41, 239}

\bibitem[\protect\citeauthoryear{{Tully}}{{Tully}}{1988}]{1988ngc..book.....T}
{Tully} R.~B.,  1988, {Nearby galaxies catalog}

\bibitem[\protect\citeauthoryear{{Tully}, {Rizzi}, {Shaya}, {Courtois},
  {Makarov}  \& {Jacobs}}{{Tully} et~al.}{2009}]{2009AJ....138..323T}
{Tully} R.~B.,  {Rizzi} L.,  {Shaya} E.~J.,  {Courtois} H.~M.,  {Makarov}
  D.~I.,   {Jacobs} B.~A.,  2009, \mn@doi [\aj] {10.1088/0004-6256/138/2/323},
  \href {http://adsabs.harvard.edu/abs/2009AJ....138..323T} {138, 323}

\bibitem[\protect\citeauthoryear{{Tully} et~al.,}{{Tully}
  et~al.}{2013}]{2013AJ....146...86T}
{Tully} R.~B.,  et~al., 2013, \mn@doi [\aj] {10.1088/0004-6256/146/4/86}, \href
  {http://cdsads.u-strasbg.fr/abs/2013AJ....146...86T} {146, 86}

\bibitem[\protect\citeauthoryear{{Vinokurov}, {Fabrika}  \&
  {Atapin}}{{Vinokurov} et~al.}{2016}]{2016arXiv160603024V}
{Vinokurov} A.,  {Fabrika} S.,   {Atapin} K.,  2016, preprint, \href
  {http://adsabs.harvard.edu/abs/2016arXiv160603024V} {} (\mn@eprint {arXiv}
  {1606.03024})

\bibitem[\protect\citeauthoryear{{Walton}, {Roberts}, {Mateos}  \&
  {Heard}}{{Walton} et~al.}{2011}]{2011MNRAS.416.1844W}
{Walton} D.~J.,  {Roberts} T.~P.,  {Mateos} S.,   {Heard} V.,  2011, \mn@doi
  [\mnras] {10.1111/j.1365-2966.2011.19154.x}, \href
  {http://adsabs.harvard.edu/abs/2011MNRAS.416.1844W} {416, 1844}

\bibitem[\protect\citeauthoryear{{Wenger} et~al.,}{{Wenger}
  et~al.}{2000}]{2000A&AS..143....9W}
{Wenger} M.,  et~al., 2000, \mn@doi [\aaps] {10.1051/aas:2000332}, \href
  {http://adsabs.harvard.edu/abs/2000A%26AS..143....9W} {143, 9}

\bibitem[\protect\citeauthoryear{{Winter}, {Mushotzky}  \& {Reynolds}}{{Winter}
  et~al.}{2006}]{2006ApJ...649..730W}
{Winter} L.~M.,  {Mushotzky} R.~F.,   {Reynolds} C.~S.,  2006, \mn@doi [\apj]
  {10.1086/506579}, \href {http://cdsads.u-strasbg.fr/abs/2006ApJ...649..730W}
  {649, 730}

\bibitem[\protect\citeauthoryear{{Wright} et~al.,}{{Wright}
  et~al.}{2010}]{2010AJ....140.1868W}
{Wright} E.~L.,  et~al., 2010, \mn@doi [\aj] {10.1088/0004-6256/140/6/1868},
  \href {http://adsabs.harvard.edu/abs/2010AJ....140.1868W} {140, 1868}

\bibitem[\protect\citeauthoryear{{Wu} et~al.,}{{Wu}
  et~al.}{2015}]{2015Natur.518..512W}
{Wu} X.-B.,  et~al., 2015, \mn@doi [\nat] {10.1038/nature14241}, \href
  {http://adsabs.harvard.edu/abs/2015Natur.518..512W} {518, 512}

\bibitem[\protect\citeauthoryear{{Zacharias}, {Finch}, {Girard}, {Henden},
  {Bartlett}, {Monet}  \& {Zacharias}}{{Zacharias}
  et~al.}{2013}]{2013AJ....145...44Z}
{Zacharias} N.,  {Finch} C.~T.,  {Girard} T.~M.,  {Henden} A.,  {Bartlett}
  J.~L.,  {Monet} D.~G.,   {Zacharias} M.~I.,  2013, \mn@doi [\aj]
  {10.1088/0004-6256/145/2/44}, \href
  {http://cdsads.u-strasbg.fr/abs/2013AJ....145...44Z} {145, 44}

\makeatother
\end{thebibliography}


\appendix

\section{Tables}

\begin{table*}
\vspace{5mm}
\begin{center}
\caption{Coordinates of the identified NIR candidate counterparts to the ULXs listed in Table 5 of H14, obtained by the `pick object' tool in {\scshape gaia}. The classification of the NIR candidate counterparts is based on their absolute magnitudes, WISE colours, spatial extent and/or visual inspection of the NIR image. Spectra has been taken from 7 sources to confirm their nature.}
\label{tab:cand-marianne}
\resizebox{\textwidth}{!}{\begin{tabular}{|llcccccc|}
\hline\hline
Galaxy & ULX name & R.A & Dec. & Position$^*$ & Apparent & Absolute & Classification\\
 & in H14 & & & uncertainty & magnitude & magnitude &\\
 & & (hh:mm:ss) & (dd:mm:ss) & ( \arcsec\ ) & (mag) & (mag) &\\ 
\hline\hline
NGC 253 & J004722-252051 & 00:47:22.60 & -25:20:51.30 & 0.78 & 17.2 $\pm$ 0.03 $\pm$ 0.5 & -10.5 $\pm$ 0.03 $\pm$ 0.5 $\pm$ 0.10 & RSG$^a$\\
NGC 925 & J022721+333500 & 02:27:21.53 & +33:35:00.70 & 0.84 & 18.7 $\pm$ 0.03 $\pm$ 0.2 & -10.6 $\pm$ 0.03 $\pm$ 0.2 $\pm$ 0.4 & RSG$^b$\\
NGC 925 & J022727+333443 & 02:27:27.56 & +33:34:43.50 & 0.84 & 20.1 $\pm$ 0.08 $\pm$ 0.2 & -9.2 $\pm$ 0.08 $\pm$ 0.2 $\pm$ 0.4 & N$^b$\\
NGC 1058 & J024323+372038 & 02:43:23.28 & +37:20:42.48 & 0.72 & 19.7 $\pm$ 0.06 $\pm$ 0.4 & -10.1 $\pm$ 0.06 $\pm$ 0.4 $\pm$ 0.4 & cRSG\\
NGC 1637 & [IWL2003 68] & 04:41:32.9 & -02:51:26.2 & 1.2 & 16.3 $\pm$ 0.005 $\pm$ 0.5 & -13.7 $\pm$ 0.005 $\pm$ 0.5 $\pm$ 0.4 & SC/AGN\\
NGC 2500 & J080157+504339 & 08:01:57.86 & +50:43:39.96 & 0.18 & 15.7 $\pm$ 0.002 $\pm$ 0.2 & -14.1 $\pm$ 0.005 $\pm$ 0.15 $\pm$ 0.4 & AGN$^c$\\
Holmberg II & Holmberg II X-1 & 08:19:28.94 & +70:42:19.71 & 0.66 & 19.30 $\pm$ 0.08 $\pm$ 0.10 & -8.35 $\pm$ 0.08 $\pm$ 0.10 $\pm$ 0.03 & cRSG$^d$\\
Holmberg I & Ho I XMMI & 09:41:30.23 & +71:12:35.63 & 0.42 & 17.81 $\pm$ 0.01 $\pm$ 0.10 & -10.14 $\pm$ 0.01 $\pm$ 0.10 $\pm$ 0.03 & AGN\\
NGC 3627 & J112018+125900 & 11:20:18.29 & +12:59:00.93 & 0.72 & 20.6 $\pm$ 1.9 $\pm$ 0.7 & -9.1 $\pm$ 1.9 $\pm$ 0.7 $\pm$ 0.4 & cRSG\\
NGC 4136 & J120922+295551 & 12:09:22.63 & +29:55:50.98 & 1.02 & 19.13 $\pm$ 0.03 $\pm$ 0.10 & -10.78 $\pm$ 0.03 $\pm$ 0.10 $\pm$ 0.4 & cRSG$^d$\\
NGC 4136 & J120922+295559 & 12:09:22.19 & +29:55:59.03 & 1.02 & 19.15 $\pm$ 0.03 $\pm$ 0.10 & -10.75 $\pm$ 0.03 $\pm$ 0.10 $\pm$ 0.4 & RSG$^b$\\
NGC 4258 & J121844+471730 & 12:18:43.9 & +47:17:31.0 & 1.5 & 17.79 $\pm$ 0.02 $\pm$ 0.10 & -11.50 $\pm$ 0.02 $\pm$ 0.1 $\pm$ 0.02 & SC\\
NGC 5194 & J132953+471040 & 13:29:53.29 & +47:10:42.60 & 0.84 & 15.72 $\pm$ 0.02 $\pm$ 0.10 & -13.88 $\pm$ 0.02 $\pm$ 0.10 $\pm$ 0.2 & SC\\
NGC 5408 & NGC 5408 X-1 & 14:03:19.68 & -41:22:58.63 & 0.78 & 20.3 $\pm$ 0.13 $\pm$ 0.2 & -8.1 $\pm$ 0.13 $\pm$ 0.2 $\pm$ 0.8 & cRSG\\
NGC 5457 & J1402+5440 & 14:04:14.24 & +54:26:02.86 & 0.69 & 19.3 $\pm$ 0.04 $\pm$ 0.2 & -9.7 $\pm$ 0.04 $\pm$ 0.2 $\pm$ 0.05 & cRSG\\
NGC 5457 & J140314+541807 & 14:03:14.39 & +54:18:07.10 & 0.45 & 17.72 $\pm$ 0.01 $\pm$ 0.05 & -11.32 $\pm$ 0.01 $\pm$ 0.05 $\pm$ 0.05 & cRSG\\
NGC 5457 & J140248+541350 & 14:02:48.15 & +54:13:50.56 & 0.36 & 18.35 $\pm$ 0.03 $\pm$ 0.10 & -10.69 $\pm$ 0.03 $\pm$ 0.10 $\pm$ 0.05 & AGN\\
\hline\hline
\multicolumn{8}{l}{{\bf Notes:} Confirmed nature by: $^a$\citet{2015MNRAS.453.3510H}, $^b$\citet{2016MNRAS.459..771H}, $^c$\citet{2013A&A...549A..81G}. $^d$: Spectra was taken by \citet{2016MNRAS.459..771H}, but yielded}\\
\multicolumn{8}{l}{no conclusion. $^*$: 99.7$\%$ uncertainty radius around the position of the NIR candidate counterpart. The abbreviations are: cRSG -- candidate RSG,}\\
\multicolumn{8}{l}{SC -- stellar cluster, N -- Nebula.}\\
\end{tabular}}
\end{center}
\end{table*}


\bsp	
\label{lastpage}
\end{document}